\documentclass[12pt]{article} 
\usepackage{dina4p} 
\usepackage{epsfig}  
\usepackage{wrapfig}
\def\Journal#1#2#3#4{{#1} {\bf #2}, #3 (#4)}

\def\NPA{{\em Nucl. Phys.} A}
\def\NPB{{\em Nucl. Phys.} B}
\def\PLB{{\em Phys. Lett.}  B}
\def\PRL{\em Phys. Rev. Lett.}
\def\PRD{{\em Phys. Rev.} D}
\def\ZPC{{\em Z. Phys.} C}
\def\lsim{\mathrel{\rlap{\lower4pt\hbox{\hskip1pt$\sim$}}
    \raise1pt\hbox{$<$}}}                
\def\gsim{\mathrel{\rlap{\lower4pt\hbox{\hskip1pt$\sim$}}
    \raise1pt\hbox{$>$}}}                

\begin{document}

\vspace*{-12mm}

\begin{flushright}
   DESY 99--122    \\
   August 1999     
\end{flushright}

\begin{center}  \begin{Large} \begin{bf}
The Physics Case for \\
Polarised Proton-Nucleon Scattering at HERA\footnote{Invited talk
  at the workshop `Polarized Protons at High Energies - Accelerator
  Challenges and Physics Opportunities, DESY Hamburg, 17-20 May
  1999.} 
   \\
  \end{bf}  \end{Large}
  \vspace*{4mm}
  \begin{large}
V. A. Korotkov$^{a,b}$ and \underline{W.-D. Nowak}$^c$
  \end{large}
\end{center}
$^a$ Deutsches~Elektronen-Synchrotron~DESY, 
     D-22603~Hamburg,~Germany \\
$^b$ IHEP, RU-142284 Protvino, Russia \\
$^c$ DESY Zeuthen, D-15735~Zeuthen, Germany 
\begin{quotation}
\noindent
{\bf Abstract:}
The physics case for a possible fixed target polarized nucleon-nucleon
collision experiment at HERA is described.
The experiment named {\it HERA--}$\vec{N}$ could be realized using
an internal polarized gas target in the HERA polarized/unpolarized
proton beam. 
A wide spectrum of nucleon spin structure problems could be
investigated and the experiment
would constitute a fixed target complement to the RHIC 
spin physics program with competitive statistical accuracy.

\end{quotation}
\section{Introduction}
For more than 4 years {\it HERMES} at DESY has been
studying the spin structure of the nucleon using a
polarized internal gas
target in HERA's polarized electron/positron storage ring. 
In a similar approach, the
installation of such a target in HERA's {\it proton} storage ring 
would open a new chapter in the study of high energy nucleon-nucleon
spin physics in a fixed-target environmnent at
$\sqrt{s}~\simeq~40$~GeV \cite{wdnTrieste,KoNoNPA}.
An internal polarized nucleon target offering 
a polarization above 80\% and no or small dilution, can be safely
operated in a high energy proton ring at high densities up to $10^{14}$
atoms/cm$^2$.
As long as the polarized target is used in
conjunction with the unpolarized proton beam, the physics scope would
be focused onto `phase~I', i.e. measurements of 
single spin asymmetries. Once polarized protons should be available,
a variety of double spin asymmetries could be investigated. These
`phase~II' measurements would then constitute an alternative --~fixed
target~-- approach to similar physics which soon will be accessible to
the collider experiments at the low end of the
RHIC energy scale ($\sqrt{s}~\simeq~50$~GeV). Altogether,
a rich spin physics program would emerge at DESY; it has been called 
{\it HERA--}$\vec{N}$ \cite{wdnTrieste} to allow for easier reference. \\
The estimate of the integrated luminosity which could be accumulated in
 the experiment is based upon realistic figures. 
For the average beam and target polarisation
$P_B = 0.6$ and $P_T = 0.8$ are assumed, respectively. A combined
trigger and reconstruction efficiency of $C \simeq 50\%$ is
anticipated. Using rather conservative values for both the average
HERA proton beam current 
($\bar{I}_B = 80 \; \mbox{mA} = 0.5 \cdot 10^{18} \; s^{-1}$)
and for the polarized target density 
($n_T = 3 \cdot 10^{13}$ atoms/cm$^2$) 
the projected integrated luminosity becomes
${\cal{L}} \cdot T =$ 240~pb$^{-1}$
when for the total running time $T$ an equivalent 
of $T = 1.6 \cdot 10^7 \;s$ 
is assumed. This corresponds to about 3 real years under
{\it present} HERA conditions. 
One may argue, however, that at the time the experiment would run
 even 500~pb$^{-1}$ {\it per year} might 
presumably become a realistic figure 
and the luminosity to be
accumulated over the lifetime of the experiment might be considerably higher.

\section{Single Spin Asymmetries}
Single (transverse) spin asymmetries in inclusive particle
production at large $p_T$ 
are forbidden in leading-twist pQCD,
reflecting the fact that single spin asymmetries are zero at the 
partonic level and that collinear parton configurations inside hadrons
do not allow for single spin dependences. The earlier naive expectation
that they should generally be zero in pQCD has 
been proven to be false.
Several models and theoretical analyses suggest possible higher
twist effects: there might be twist-3 dynamical contributions
(hard scattering higher twists); there might also be 
intrinsic $k_\perp$ effects, both in the quark fragmentation process 
\cite{collins} (Collins effect)
and in the quark distribution functions \cite{sivers} (Sivers effect). 
It has been shown that models based on both of the latter effects
\cite{anselmino98,anselmino99} are capable to describe the existing 
experimental data.

Qiu and Sterman recently presented a detailed discussion of
single spin asymmetries in pion production \cite{QiuSt2} in the 
context of twist-3 matrix elements.
At moderate $x_F$, a rather mild decrease with $p_T$ is predicted
already for transverse momenta above 2 GeV; this is the region where
the perturbative calculations are expected to be reliable. 

There exists a variety of alternative approaches to the above
mentioned twist-3 based explanations of single spin asymmetries. As an
example, one model derives the asymmetry from the orbital momenta of
the current quarks within the constituent quark \cite{troshin}. Sign
and size of the asymmetry are then proportional to the polarization of
the constituent quark in the polarized nucleon. Above 2 GeV an
$\cal{O}$(0.3) and almost $p_T$-independent asymmetry is predicted
even at the highest RHIC energy, $\sqrt{s} = 500$ GeV. In a totally
different approach, based on the contribution of instantons to the
hadron production processes, the single spin asymmetry is predicted
\cite{kochelev} in qualitative agreement with existing experimental data.

In the following the capabilities of {\it HERA--}$\vec N$ are discussed
to investigate single spin asymmetries with a transversely polarized
target in the unpolarized HERA proton beam (`phase I'). It is worth
noting that once polarization of the proton beam should have been
achieved (`phase II'), single spin asymmetries may be measured with
significantly higher precision using an unpolarized target whose
density is only limited by beam life time deterioration. 

\begin{wrapfigure}{r}{8.0cm}
\vspace*{-12mm}
\begin{center}
\epsfig{file=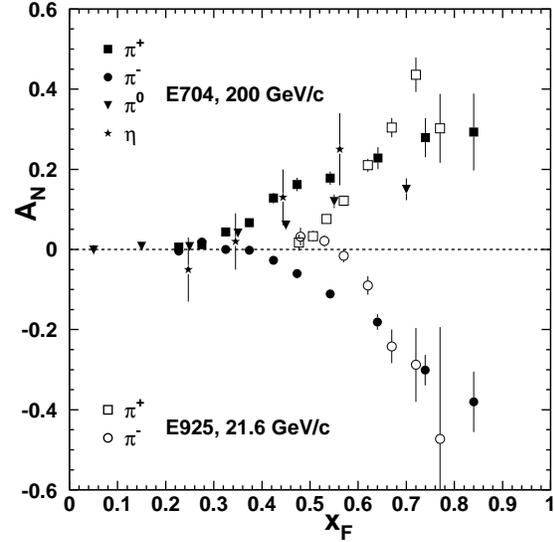,width=8.0cm}
\caption{
\it Single spin asymmetry in inclusive pion and eta production
         as measured by the {\it E704}  \protect \cite{704pi}
          and {\it E925} \protect \cite{exp925} Collaborations.
}
\label{e704data}
\end{center}
\end{wrapfigure}

%
{\bf Inclusive pion and kaon production.}
The reaction  $p^{\uparrow} p \rightarrow \pi^{0\pm}X$
exhibits surprisingly large single spin asymmetries, as was
measured a few years ago by the {\it E704} Collaboration at Fermilab
using a 200 GeV transversely polarized proton beam \cite{704pi}. For
any kind of pions the asymmetry $A_N$ shows a considerable rise above
$x_F > 0.3$, i.e. in the fragmentation region of the polarized nucleon
(see fig.~\ref{e704data}).  
It is positive for both $\pi^+$ and $\pi^0$ mesons, while it has 
the opposite sign for $\pi^-$ mesons. The same tendency was observed
recently by the {\it E925} Collaboration at BNL scattering a 21.6 GeV
proton beam on a carbon target \cite{exp925}. Their data are shown in
fig.~\ref{e704data} together with the older {\it E704} data. At large
$x_F$ ($\gsim$ 0.6) the measured asymmetries are compatible in both
experiments, while at $x_F \lsim$ 0.5 the asymmetries measured by {\it
  E925} are compatible with zero. This difference 
may be induced not only by the different beam energies but also 
by different acceptances in $p_T$ and different target nuclei used.
The charged pion data of {\it E704} taken in the range 
$0.2 < p_T < 2$~GeV were split into two samples at $p_T$~=~0.7~ GeV/c;
the observed rise is stronger for the high $p_T$ sample.
New results on the asymmetry in $\eta$ meson production were presented 
recently by {\it E704} \cite{704eta}. 
The asymmetry is positive and the behaviour is
compatible with the one observed in $\pi^0$ and $\pi^+$ production. 
A variety of asymmetry measurements exists in
inclusive particle production at smaller energies (see e.g. a recent
review in ref.~\cite{bravar}).

\begin{figure}[htb]
\centering
\begin{minipage}[c]{5.5cm}
\centering
\epsfig{file=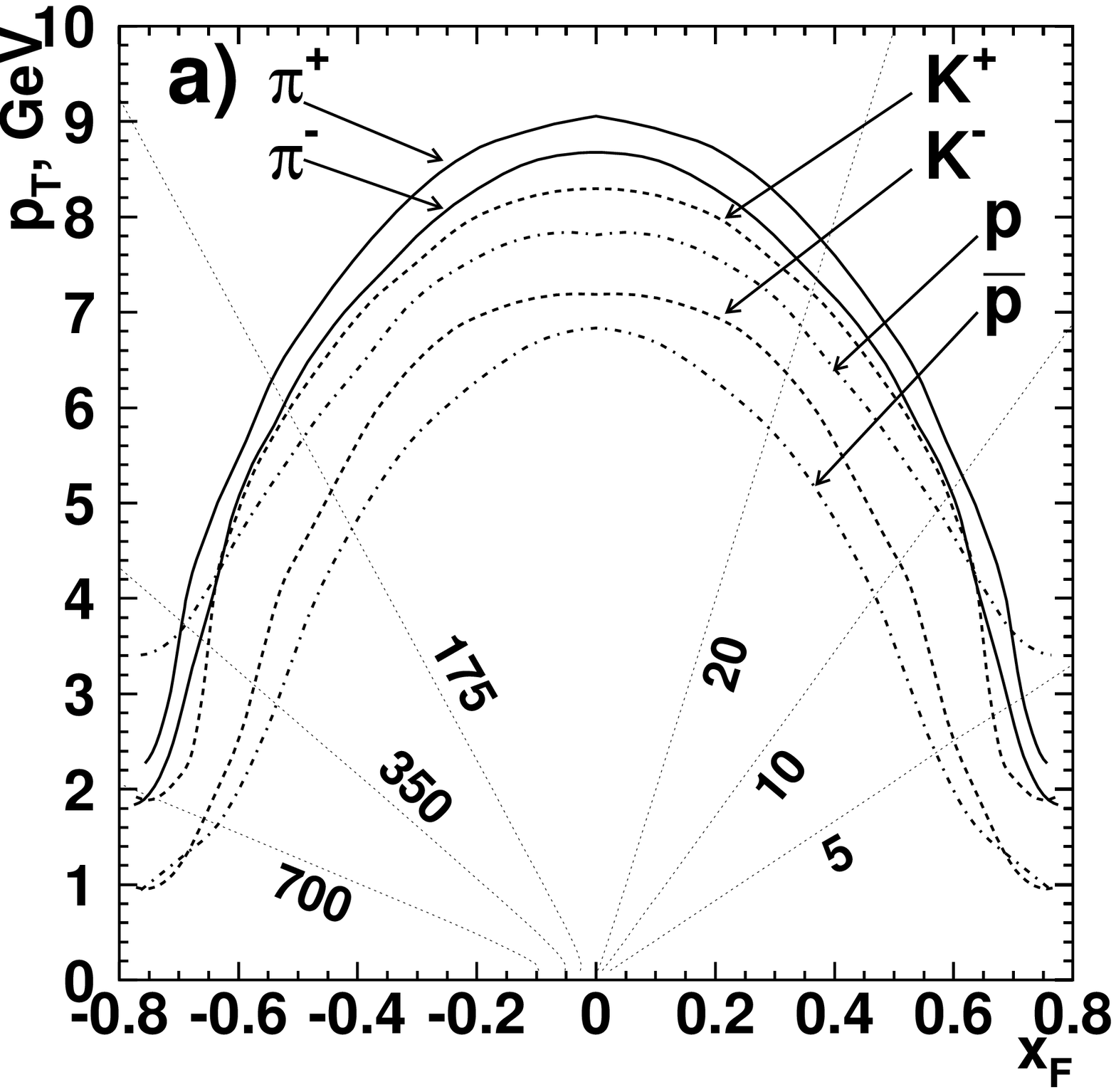,width=5.5cm}
\end{minipage}
\begin{minipage}[c]{5.5cm}
\centering
\epsfig{file=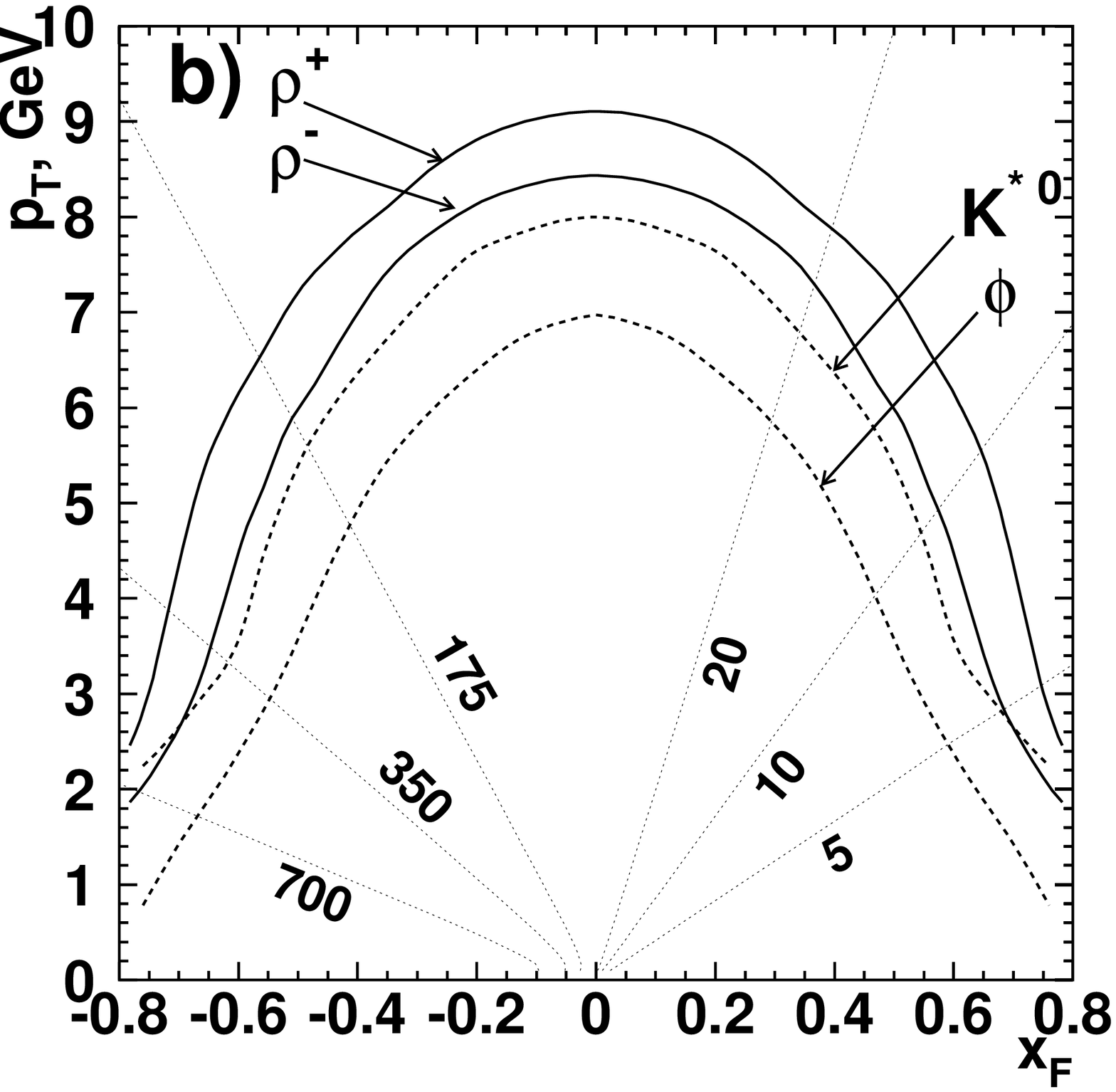,width=5.5cm}
\end{minipage}
\begin{minipage}[c]{5.5cm}
\centering
\epsfig{file=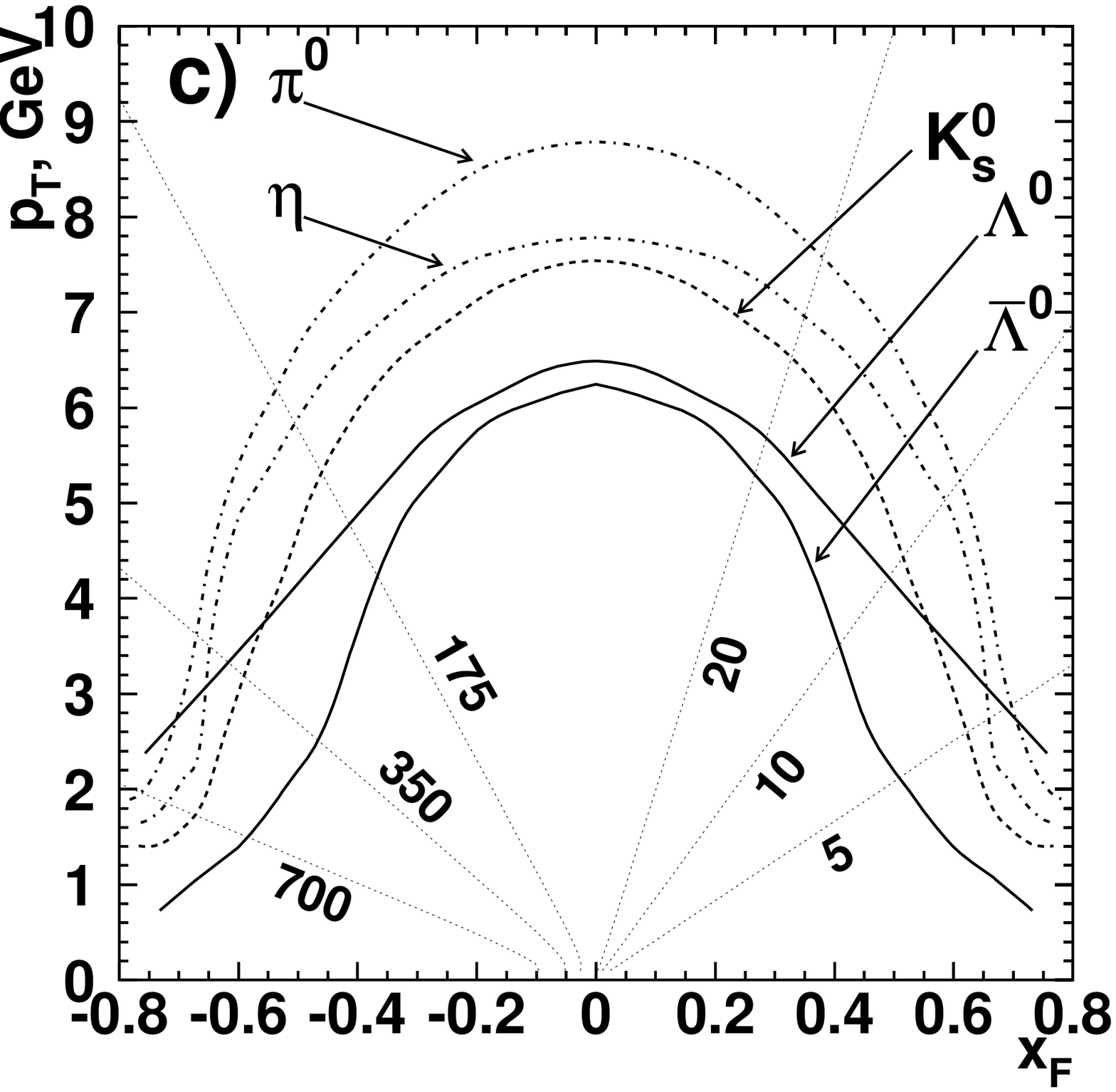,width=5.5cm}
\end{minipage}
\caption{
\it \it Contours of the asymmetry sensitivity level 
     for inclusive production of different
     particles in the $(x_F, p_T)$ plane; inside the contour the
     sensitivity level is $\delta A_N \leq 0.05$. Only the following decay
     modes are taken into account: $K^{*0} \rightarrow K^+ \pi^-$,
     $\phi \rightarrow K^+ K^-$, $\eta \rightarrow \gamma \gamma$,
     $K^{0}_s \rightarrow \pi^+ \pi^-$, $ \Lambda^0 \rightarrow p
     \pi^-$, $ \bar{\Lambda}^0 \rightarrow \bar p \pi^+$. Lines of
     constant laboratory angles of the particles are shown and marked
     with their values in units of mrad. }
\label{ssaAllPart}
\end{figure}

The $p_T$ values accessible with {\it HERA--}$\vec{N}$ would be
significantly larger than those reached in all experiments performed
up to now. The sensitivity 
$\delta A_N$ of the asymmetry measurement in inclusive production of
different particles at {\it HERA--}$\vec{N}$ was calculated
\cite{KoNoZtn97} 
using the inclusive differential cross-sections obtained with the
Monte-Carlo program PYTHIA~5.6 \cite{pythia}. The results are shown in
fig.~\ref{ssaAllPart} in the ($x_F$, $p_T$) plane as contours
characterizing the sensitivity level $\delta A_N = 0.05$ in bins of
$\Delta p_L \times \Delta p_T = 2 \times 2 $ (GeV/c)$^2$. For produced
particles lines of constant polar angle in the laboratory system are
shown.

Experimentally, it is not a simple task to measure single spin
asymmetries in the fragmentation region of the polarized nucleon in a
fixed target experiment at 820~GeV. This region lies either at
negative $x_F$, i.e. at very
large laboratory angles (a few tens of degrees), if a combination of
polarized target and unpolarized beam is used, or at positive $x_F$,
i.e. at very small
angles (a few mrad) for the other combination, unpolarized target and 
polarized beam (cf. fig.~\ref{ssaAllPart}). The question, how close to
the HERA proton beam particles can be measured, deserves a special
study. 

As can be seen from fig.~\ref{ssaAllPart}, the combined $p_T$
dependence of all involved 
effects can be measured with
good statistical accuracy ($\delta A_N \leq 0.05$) up to transverse
momenta of about 8$\div$10~GeV/c in the central region $|x_F| < 0.2$
and up to 5$\div$6~GeV/c in the target or beam fragmentation region,
respectively, depending on which nucleon is polarized. Hence the
$p_T$ range of a few GeV, in which higher twist effects are expected
to be essential, would be well covered. \\

\begin{figure}
\begin{center}
\epsfig{file=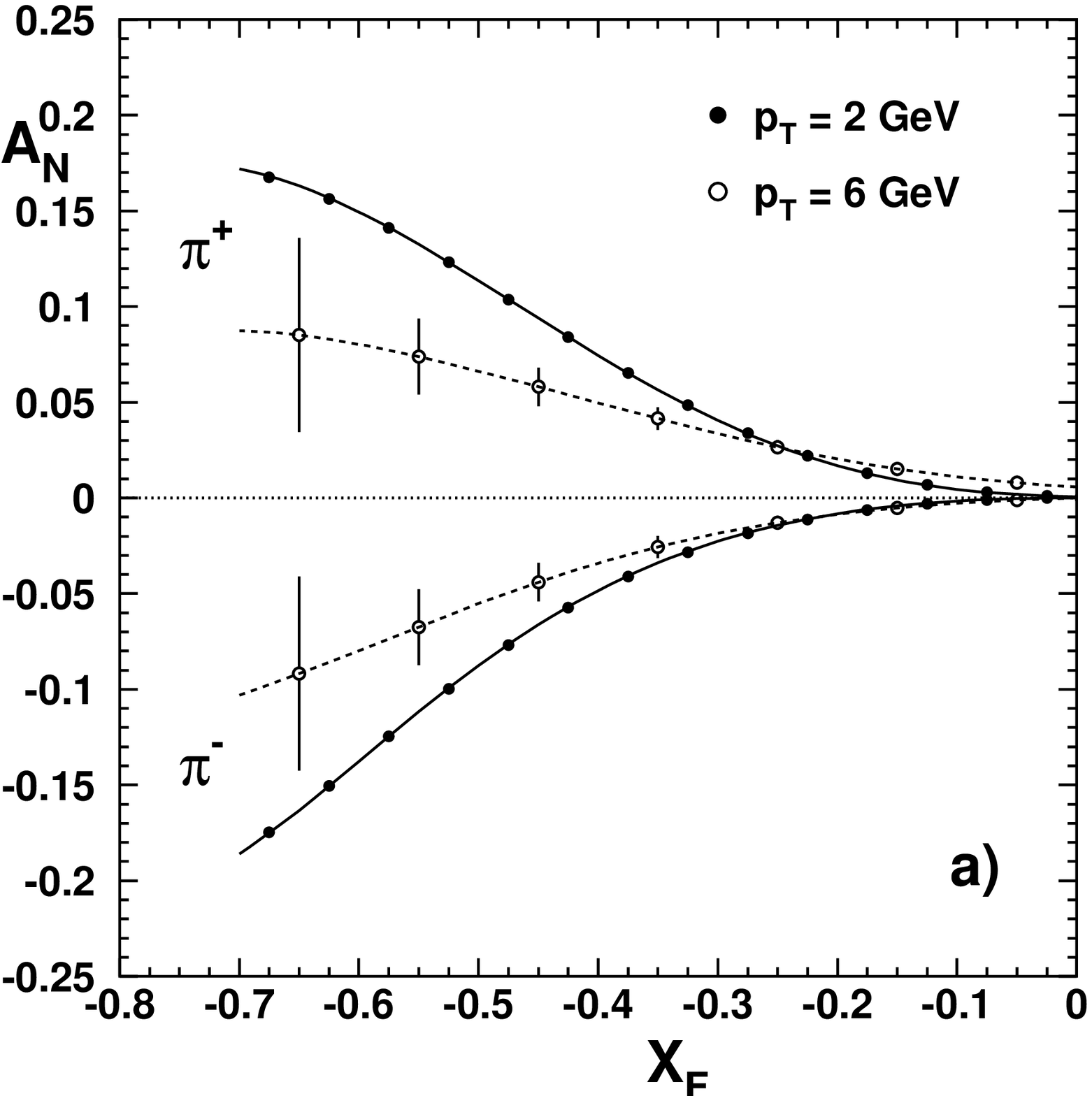,width=8.0cm}
\epsfig{file=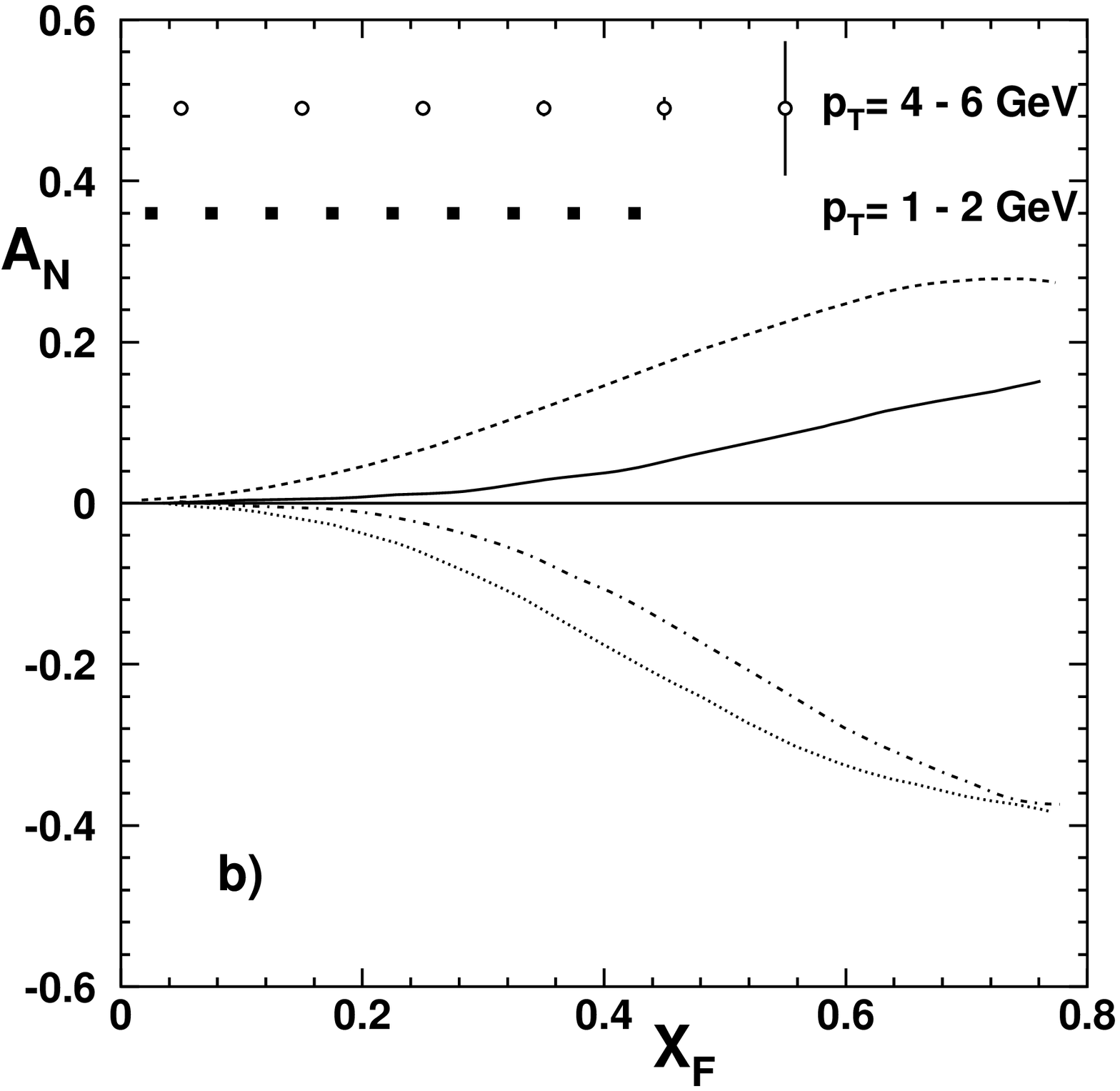,width=8.0cm}
\caption{
\it (a) Capability of {\it HERA--}$\vec{N}$ to prove the $p_T$ dependence of
    $A_N$ in charged pion production, as predicted by the model of 
    ref.~\cite{anselmino98}.
    (b) Capability of {\it HERA--}$\vec{N}$ to measure a $p_T - x_F$ 
    dependence of the asymmetry in inclusive $K_s^0$ production.
    The curves correspond to different sets of kaon fragmentation 
    function (see ref.~\cite{anselmino98}).  } 
\label{asymurgia}
\end{center}
\end{figure}

The capability of {\it HERA--}$\vec{N}$
to prove or disprove a predicted $p_T$ dependence in the fragmentation
region of the polarized nucleon is shown in fig.~\ref{asymurgia}a for
inclusive pion production. The
theoretical curve is based on the assumption of a non--zero quark
distribution 
analysing power according to ref.~\cite{anselmino98}. Note
that a somewhat weaker increase with $x_F$ is obtained for the
predicted asymmetry value if the model of ref.~\cite{QiuSt2} was
used. The curves and the projected statistical errors in
fig.~\ref{asymurgia}a are drawn for the combination of unpolarized
proton beam and polarized target.

Fig.~\ref{asymurgia}b shows the capability of {\it HERA--}$\vec{N}$ 
to measure the $p_T$ and $x_F$ dependence of the asymmetry in
inclusive $K_s^0$ production,
calculated for the case of a polarized beam and an unpolarized target
and assuming the $K_S^0$ mesons to be accepted in the range 5 to
170~mrad. The theoretical predictions are taken from
ref.~\cite{anselmino98}. It should be noted that the existing sets of
kaon fragmentation functions essentially offer two choices. On the one
hand, if the main contribution to the large $x_F$ production of kaons
is coming from 
partonic processes involving only valence quarks from the initial
nucleon and the final meson, the single spin asymmetries predicted by 
the model of ref.~\cite{anselmino98} are quite similar to the pion case:  
$A_N(K^+) \sim A_N(\pi^+)$;  $A_N(K^0) \sim A_N(\pi^-)$;
$A_N(K^-) \sim A_N(\bar{K^0}) \sim 0$. 
On the other hand, if the used set of kaon fragmentation functions
enhances the role of the sea quarks, the situation for charged kaons
remains essentially unchanged whereas quite different predictions are
obtained for $K_S^0$s. Hence the measurement of single spin
asymmetries in inclusive neutral kaon production may offer,
as a by-product, a tool to possibly discriminate between different 
sets of kaon fragmentation functions.

{\bf Inclusive direct photon production.}
The reaction $p p^{\uparrow} \to \gamma X$ proceeds without
fragmentation, i.e. 
this process measures
a combination of initial $k_{\perp}$ effects and hard scattering twist--3
processes. The first and only results up to now were obtained by the
{\it E704} Collaboration \cite{Phot704} showing an asymmetry
compatible with zero within large errors for $2.5 < p_T <3.1$~GeV/c in
the central region $ | x_F | \lsim 0.15$.   

\begin{wrapfigure}{l}{8.0cm}
\vspace*{-15mm}
\begin{center}
\epsfig{file=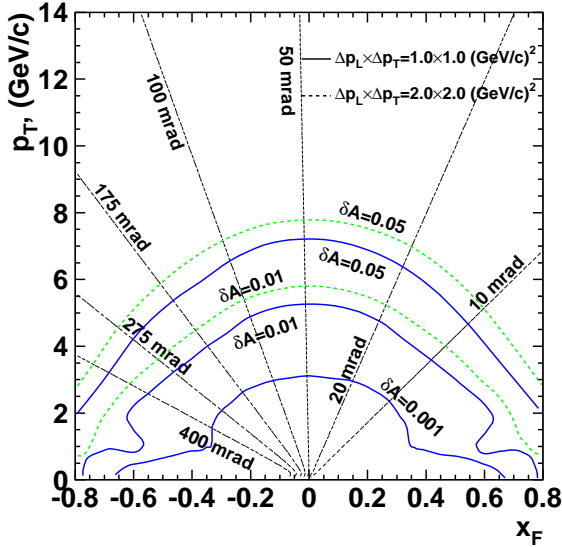,width=8cm}
\vspace*{-5mm}
\caption
{\it Asymmetry sensitivity levels for photon
       production in the ($p_T,~x_F$) plane. 
       Laboratory angles of the photons are
       shown.}
\label{gamxfpt}
\end{center}
\end{wrapfigure}

The experimental sensitivity of {\it HERA--}$\vec N$ (see
fig.~\ref{gamxfpt}) was determined using cross-section calculations of
the two dominant hard subprocesses,
i.e. gluon--Compton scattering
($qg \rightarrow \gamma q$) and quark--antiquark annihilation
($q \bar q \rightarrow \gamma g$). Background photons that
originate mainly from $\pi^0$ and $\eta$ decays were taken into
account, as well.
It turns out that a good sensitivity (about 0.05)
can be maintained up to $p_T \leq$ 8 GeV/c in the central region.
For increasing transverse momentum 
the background photons are becoming less essential; hence in the central
region it is expected to be possible to detect a clear dependence
of the direct photon single spin asymmetry on $p_T$. The situation is less
favourable concerning its $x_F$ dependence. The theoretical
predictions \cite{anselmino98} show a rather shallow increase 
of $A_N$ with $x_F$. \\

\vspace*{-5mm}

{\bf Inclusive vector meson production.}
The study of polarization asymmetries in inclusive vector meson
production is especially attractive as these particles are produced  
'more directly' in comparison to pions and kaons which are mainly
decay products of heavier particles.
Comparing asymmetries in vector and
pseudoscalar meson production can provide information on the magnitude
of the asymmetry in quark scattering \cite{czyz}. If the asymmetry is
generated only during the fragmentation of polarized quarks, the
asymmetry of $\rho$ mesons is expected to be opposite in sign to that
of pions, $R_{\rho / \pi} = A_N^\rho / A_N^\pi \simeq -{1 \over 3}$.
On the contrary, if the quark scattering asymmetry were the dominating
one, the asymmetries of pseudoscalar and vector mesons would not
differ substantially.

The statistical sensitivity of {\it HERA--}$\vec{N}$ for measuring
single spin 
asymmetries in inclusive production of $\rho$, $K^{*0}$, and $\phi$
vector mesons can be seen from fig.~\ref{ssaAllPart}b. The sensitivity
for $\rho$ production is at a level comparable to that for pions
(fig.~\ref{ssaAllPart}a, while for $K^{*0}$ and $\phi$ mesons the
reachable $p_T$ values are lower. On the other hand, a study of the
asymmetry in $K^{*0}$ and $\phi$ production using the decay channels
$K^{*0} \rightarrow K^{\pm} \pi^{\mp}$ and $\phi \rightarrow K^+ K^-$ 
could be easier since the level of the expected combinatorial
background is smaller. \\
It is worth noting that the asymmetry in $\phi$ meson production could
be useful for a study of the strange quark polarization in a nucleon
\cite{troshin}.

{\bf Inclusive Lambda production.}
A sizeable asymmetry in the inclusive production of $\Lambda^0$ and
$\bar{\Lambda}^0$ hyperons would allow to study the asymmetry in their
production up to $p_T$ of about 5 to 6~GeV/c, as can be seen from
fig.~\ref{ssaAllPart}c. The measurement of the final state  
$\Lambda$ polarization via its decay would allow to study the
polarization spin transfer coefficient, $D_{NN}$. A recent study by
{\it E704} \cite{e704lambda} at moderate values of $p_T$ (0.1$\div$1.5
GeV/c) showed a sizeable (up to 30\%) spin transfer from the incident
polarized proton to the outgoing $\Lambda^0$.   

\newpage

%
{\bf Drell-Yan production.}
The single spin asymmetry in the reaction 
$p~+~p^{\uparrow}\rightarrow~ l \bar l~+~X$ was calculated at
{\it HERA--}$\vec{N}$ energy and small transverse momenta in the
framework of twist-3 pQCD \cite{hammon,BoerTer}. The resulting
asymmetry does not exceed 0.02 in size.
It appears to be too small for an experimental verification
given the statistical significance for 240~pb$^{-1}$ that
can be seen from fig.~\ref{dygehr}.  \\

%
{\bf Inclusive {\boldmath $J/\psi$} production.}
The single spin asymmetry in inclusive J$/\psi$ production was
calculated in the framework of the colour singlet model.
The calculations at {\it HERA--}$\vec{N}$ energy 
\cite{desy96-04} show an asymmetry less than 0.01 in the region
$|x_F|~<~0.6$, i.e. the effect is practically unobservable.  \\

\begin{figure}[hbt]
\begin{center}
\epsfig{file=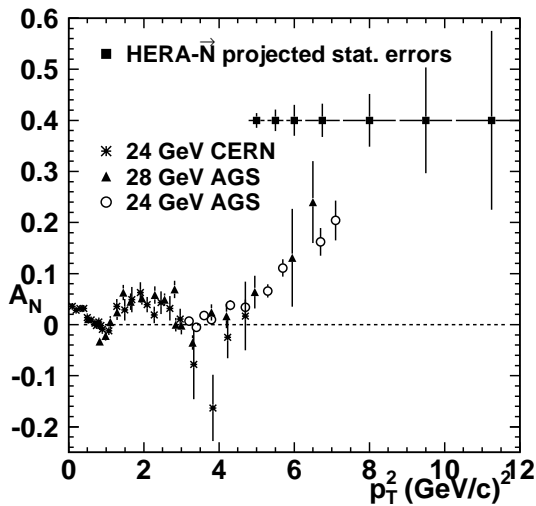,width=8cm}
\caption
{\it Single spin asymmetry in polarized proton-proton elastic
  scattering as a function of $p^2_T$.}
\label{ppelastic}
\end{center}
\end{figure}

%
{\bf Proton-proton elastic scattering.}
Large spin effects in $p~+~p^{\uparrow}\rightarrow~p~+~p$ have been
disco\-vered many years ago.
The single spin asymmetry $A_N$ was found to be significantly 
different from zero as it is 
shown in fig.~\ref{ppelastic} in conjunction 
with the projected {\it HERA--}$\vec N$ statistical errors.  
At {\it HERA--}$\vec N$ energy the detection of the recoil proton with
{\it squared} transverse momenta between 5 and 12 (GeV/c)$^2$ requires
a very large angular acceptance (up to 40 degrees) \cite{desy96-04}. 
The forward protons 
for the same interval in $p^2_T$ have laboratory angles of the order
of a few milliradians and require a dedicated forward detector very 
close to the beam pipe. 

The transverse single spin asymmetry $A_N$ in elastic $pp$ scattering 
at {\it HERA--}$\vec{N}$ and RHIC energies has been calculated in a
dynamical model that leads to spin-dependent pomeron couplings
\cite{golosk}. The predicted asymmetry is about 0.1 for 
$p_{T}^2$ values between 4 and 5 (GeV/c)$^2$ with a projected
statistical error of $0.01~\div~0.02$ for {\it HERA--}$\vec{N}$ (cf. 
fig.~\ref{ppelastic}), i.e. already with 240~pb$^{-1}$ 
a significant measurement of the asymmetry $A_N$ can be performed
to test the spin dependence of elastic $pp$ scattering at high
energies provided the necessary special detector components are
available. 

\newpage

%
\section{The Polarized Gluon Density {\boldmath $\Delta G(x)$}}
At present, the experimental information on the
polarized gluon density in the nucleon, $\Delta G(x)$, and its
moments is completely insufficient.
The only direct information comes from a recent measurement of the
HERMES Collaboration \cite{dghermes}, who found the ratio of $\Delta G
/ G$ to be positive ($\Delta G / G = 0.41 \pm 0.17 (stat.)$) at 
$x_g \sim 0.17$. However, the relatively small available c.m.s. energy of
about 7 GeV implies theoretical uncertainties that potentially are large.
In this situation new experiments are required to
accomplish both direct and indirect measurements of $\Delta G(x)$. \\

\subsection{${{\Delta G}\over{G}} (x)$ from Inclusive Processes}

{\bf Direct photon production.}
The cross section for high $p_T$ direct photon production
in $NN$ interactions is dominated by the quark--gluon Compton
subprocess, $q(x_1) + g(x_2) \rightarrow \gamma + q$; the
additional quark--antiquark annihilation subprocess is assumed to be
suppressed because of the lower density of antiquarks compared to
gluons.  \\ 
The asymmetry 
can be described by the following partonic formula
for the quark--gluon Compton subprocess: 
\begin{eqnarray}
\label{gammall}
 A_{LL}^{\; \gamma} = 
          {{ \sum_f \: e^2_f \: \Delta q_f(x_1) \: \Delta
           G(x_2) \: d \, \Delta \hat \sigma(x_1,x_2)+[x_1 \leftrightarrow x_2]}
        \over
          {\sum_f \: e^2_f \: q_f(x_1) \: G(x_2) \:
          d \, \hat \sigma(x_1,x_2)+[x_1 \leftrightarrow x_2]}}.
\end{eqnarray}
Here $x_1$ and $x_2$ are the fractional momenta of the two incoming
partons in the subprocess.
As can be seen, the asymmetry $A_{LL}^{\; \gamma}$ is
directly sensitive to the polarized gluon distribution. 

\begin{figure}
\begin{center}
\vspace*{-10mm}
\epsfig{file=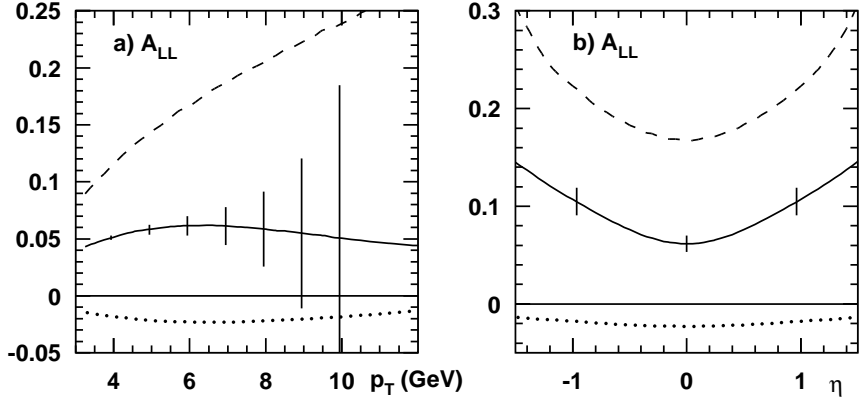,width=13cm}
\vspace*{-8mm}
\caption{
\it Double spin asymmetry in inclusive photon production
displayed vs. a $p_T$ and b $\eta$ for different polarized gluon
distribution functions (see ref. \cite{gor1}), shown in conjunction
with the projected statistical sensitivity of {\it HERA--}$\vec{N}$.}
\label{gamvogel}
\end{center}
\end{figure}

In a recent study \cite{gor1} inclusive photon production
at {\it HERA--}$\vec{N}$ was investigated. Based upon a NLO calculation,
firm predictions were obtained for $A_{LL}^{\gamma}$ including an 
assessment of the theoretical uncertainties; the latter turned out
to be of rather moderate size. Using three different assumptions for
the polarized gluon distribution corresponding predictions for the
asymmetry $A_{LL}^{\gamma}$ were calculated. Their dependence on $p_T$ and
pseudorapidity $\eta$ is shown in fig.'s \ref{gamvogel}a and
\ref{gamvogel}b in conjunction with the projected statistical
uncertainty of {\it HERA--}$\vec{N}$. As can be seen, there is
sufficient statistical accuracy over a wide kinematical region to
discriminate between different polarized gluon distribution functions. 

However, at present the theoretical interpretation of direct photon 
production at high energy is not as clear as it might appear. 
A standard NLO QCD description fails to describe the cross section 
from the recent Fermilab data of {\it E706} and requires the introduction of 
unexpectedly large intrinsic $k_\perp$ effects \cite{706QCD}. It is
likely, but still has to be proven, that this unpleasant feature will
drop out in spin asymmetries.

%
{\bf Inclusive $J/\psi$ production.}
Because of the relatively large quark mass ($m_c \gg \Lambda_{QCD}$)
the $c \bar c$ production processes occur at small
distances and the subprocess level cross sections as well as the
expected asymmetries can be calculated perturbatively. 
The underlying mechanism at the parton level is gluon-gluon fusion,
$g(x_1) + g(x_2) \rightarrow (c \bar c) + g$. 
The asymmetry can be written as
\begin{eqnarray}
\label{jpsill}
 A_{LL}^{J/\psi} = 
          {{ \sum_f \: e^2_f \: \Delta G(x_1) \: \Delta
           G(x_2) \: d \, \Delta \hat \sigma(x_1,x_2)+[x_1
           \leftrightarrow x_2]}
        \over
          {\sum_f \: e^2_f \: G(x_1) \: G(x_2) \:
          d \, \hat \sigma(x_1,x_2)+[x_1 \leftrightarrow x_2]}},
\end{eqnarray}
i.e. the asymmetry $A_{LL}^{J/\psi}$ is sensitive to the square
of the polarized gluon distribution.  

Nevertheless, also this channel is not free of theoretical problems.
The knowledge of the production mechanism is a necessary pre-requisite
for the extraction of $\Delta G$.
Experimental studies of the unpolarized production of $J/\psi$
in various reactions have shown that two possible mechanisms
should be considered to describe the existing data. The $J/\psi$
can be produced directly through the color singlet mechanism (CSM)
or in a two-step procedure through the color octet mechanism (COM).
However, both of them are not able to describe the available data
in a consistent way.

\begin{figure}
\centering
\begin{minipage}[c]{7cm}
\centering
\epsfig{file=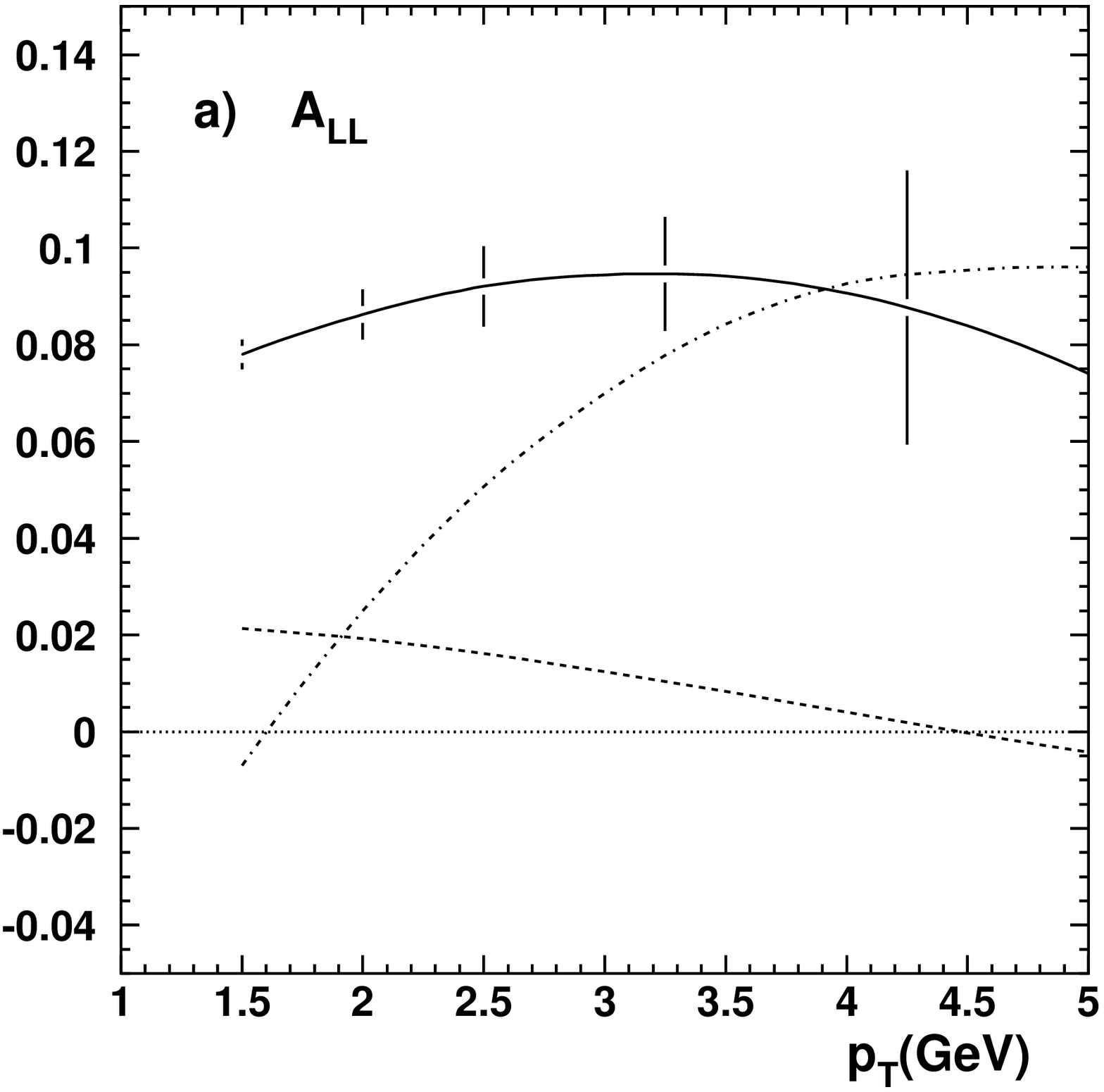,width=6.5cm}
\end{minipage}
\begin{minipage}[c]{7cm}
\centering
\epsfig{file=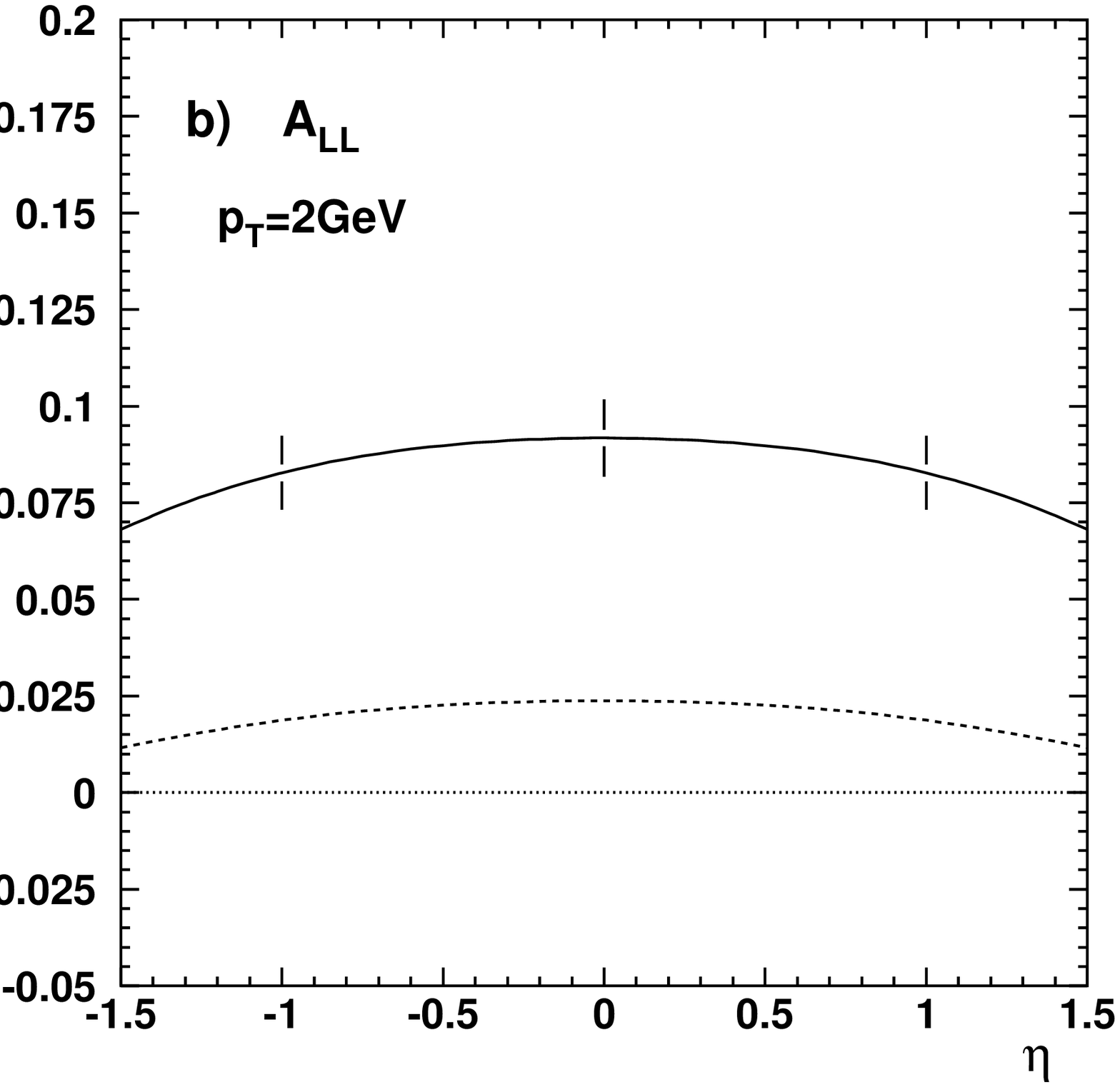,width=6.5cm}
\end{minipage}
\caption{
\it Double spin asymmetry in inclusive $J/\psi$ production
displayed vs. (a) $p_T$ 
for NLO set~A (full line) and set~B (dashed line) of
ref. \protect \cite{GSnew}; (b) $\eta$ for NLO  set~A (full line) 
and set~C (dashed line) of
ref. \protect \cite{GSnew}, shown in conjunction with the
projected statistical sensitivity of {\it HERA--}$\vec{N}$.
The dash-dotted line in (a)
represents the color singlet contribution to the asymmetry.}
\label{jpsiavto}
\end{figure}

Recently, the longitudinal double spin asymmetry in inclusive $J/\psi$
hadroproduction was calculated \cite{tertkabl} taking into account both the
color singlet and the color octet states of the $c \bar c$-pair. 
In fig.'s \ref{jpsiavto}a and \ref{jpsiavto}b the expected asymmetry
is presented versus $p_T$ and pseudorapidity $\eta$, calculated
utilizing NLO polarized  gluon distributions from ref.\cite{GSnew}.
Apparently
a very good discrimination between both sets is possible over the
whole kinematical range of {\it HERA--}$\vec{N}$; both electron and
muon decay channels of the $J/\psi$ are included.  
It is interesting to observe from fig.~\ref{jpsiavto}a that for
$p_T \gsim 3$~GeV the asymmetry originating from the total CSM
contribution is about as large as
the asymmetry caused by the COM. 
More details can be found in ref.~\cite{97-224}.

\begin{wrapfigure}{r}{8.0cm}
\vspace*{-10mm}
\begin{center}
\epsfig{file=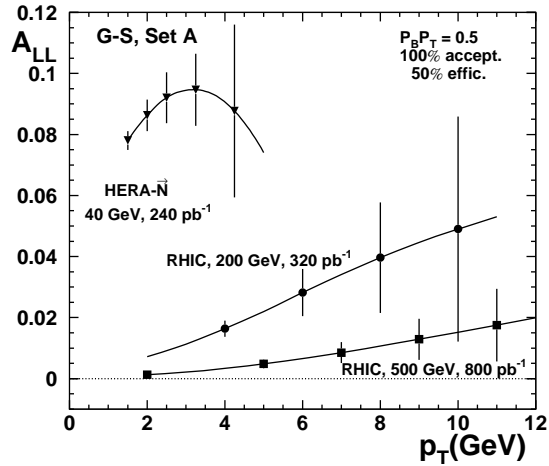,width=8cm}
\caption{
\it Transverse momentum
dependence of the expected asymmetries and projected statistical
errors for $J/\psi$ production at {\it HERA--}$\vec{N}$ and for two
different 
energies at RHIC, calculated with the NLO set~A of ref.~\cite{GSnew}. } 
\label{jpsirhic}
\end{center}
\end{wrapfigure}

For comparison the expected double spin asymmetries for inclusive
$J/\psi$ production at RHIC energies were calculated \cite{KNT-Dubna}. 
In fig.~\ref{jpsirhic} predictions at {\it HERA--}$\vec{N}$ and two
different RHIC energies are shown in conjunction with the projected
statistical uncertainties. In the
statistically accessible $p_T$ interval the asymmetry ranges between
0.08 and 0.09 at {\it HERA--}$\vec{N}$, but does not exceed 0.04
at RHIC energies. Comparing both ranges and taking into account
unavoidable limitations by systematic errors it is likely that the
fixed target experiment might accomplish a more significant
measurement of the charmonium production asymmetry. 
Additionally, the background to the $J/\psi$ signal from $b$-hadron 
decays is expected to be essential at RHIC energies. \\

{\bf Inclusive $\chi_{c1,c2}$ production.}
The two $P$-wave charmonium states $\chi_{c1}(3510), \ \chi_{c2}(3556)$
have rather large radiative $\chi_{c1,c2} \rightarrow J/\psi + \gamma$
branching ratios.
Hence optimizing the {\it HERA--}$\vec{N}$ apparatus
for both $J/\psi$ and photon detection may deliver good conditions for 
$\chi_{c1,c2}$ detection, as well. A separation of both states requires
a very good effective mass resolution of the spectrometer lying between 
10 and 20 MeV which seems not excluded at the given stage of knowledge
about a possible future apparatus. 

The above discussion of the theoretical uncertainties in the description 
of $J/\psi$ production fully applies also to $P$-wave charmonium production.

\begin{figure}
\centering
\begin{minipage}[c]{8cm}
\centering
\epsfig{file=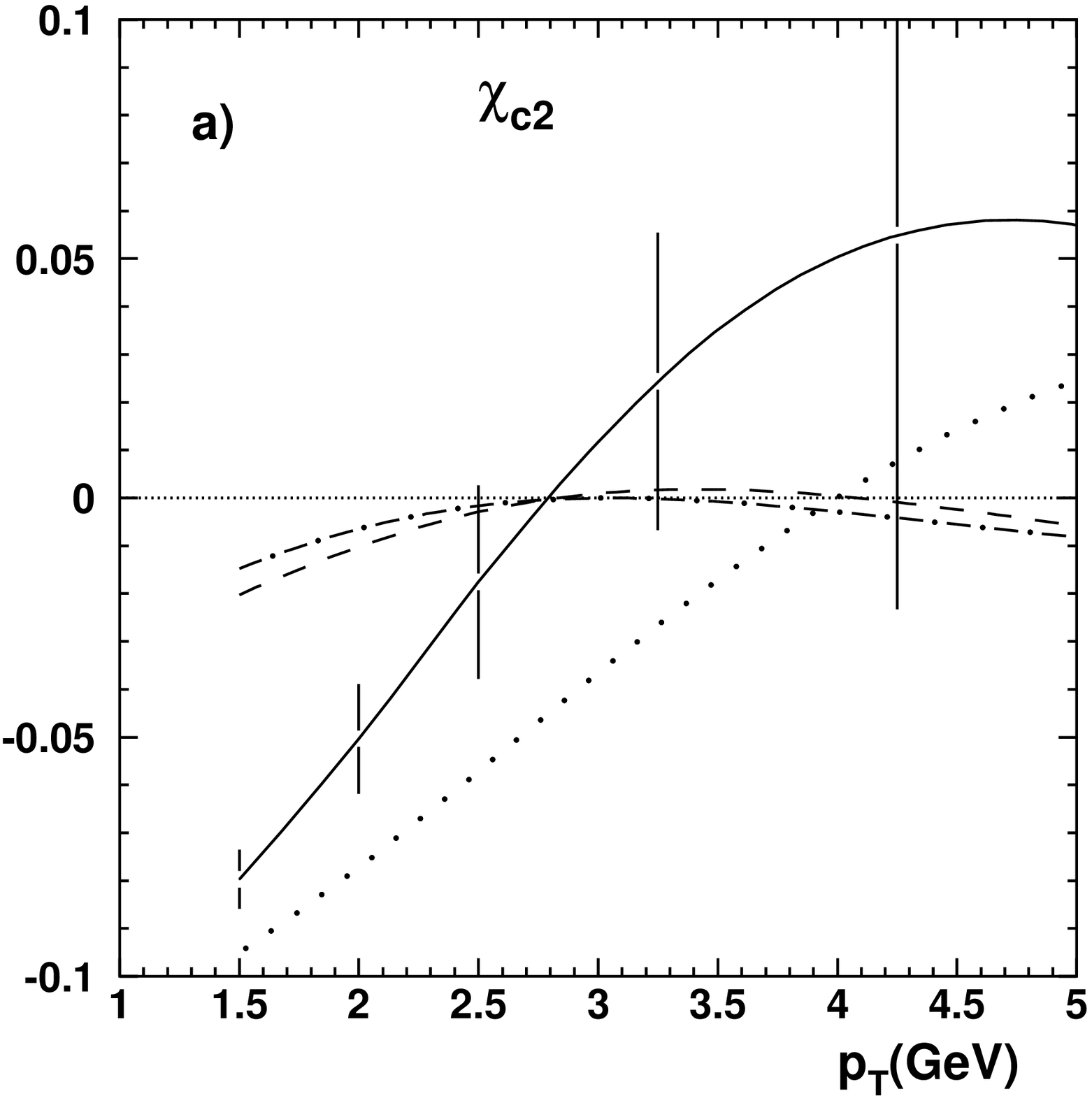,width=6cm}
\end{minipage}
\begin{minipage}[c]{8cm}
\centering
\epsfig{file=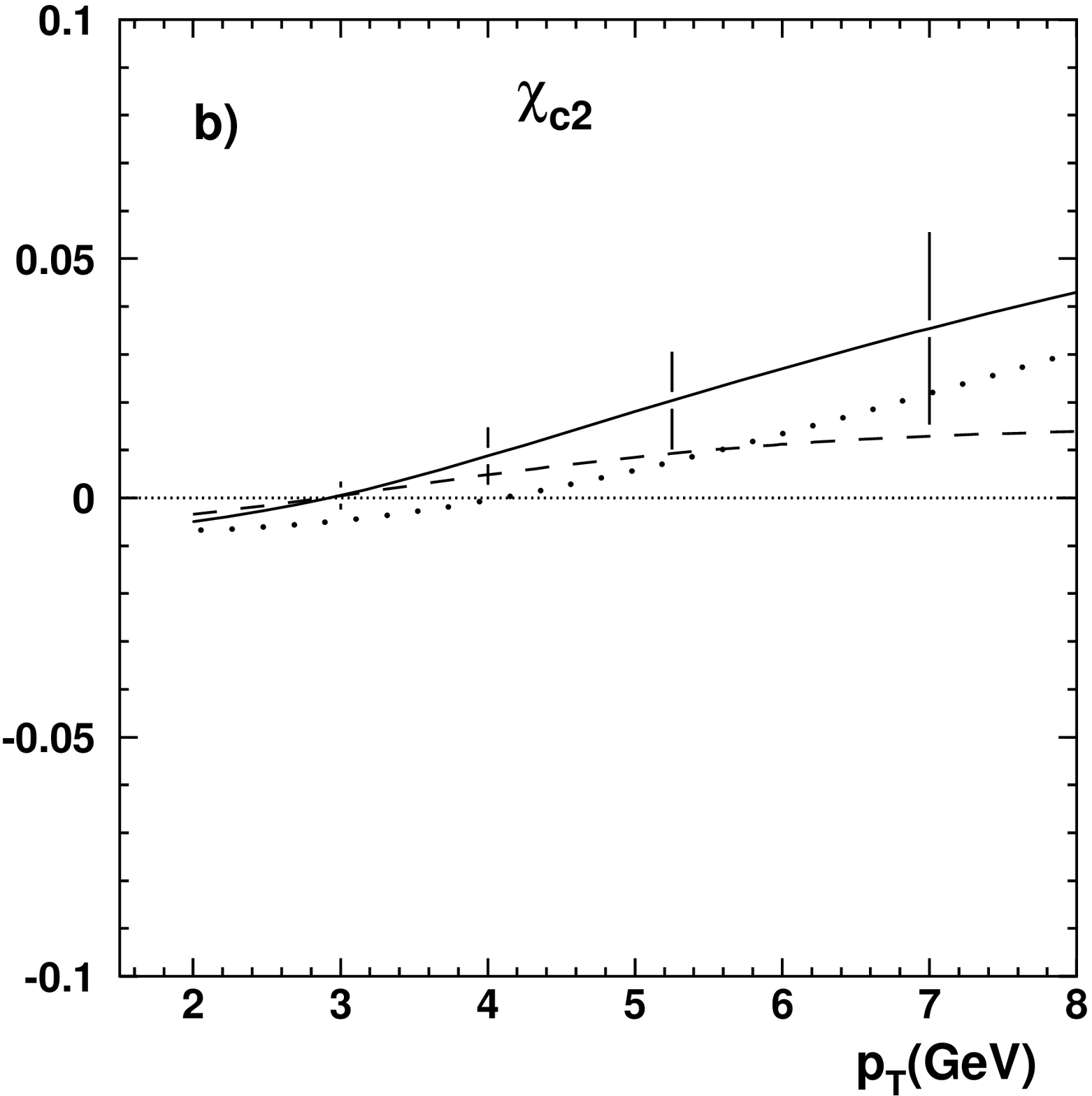,width=6cm}
\end{minipage}
\caption{
\it Double spin asymmetry in inclusive $\chi_{c2}$ production
as a function of transverse momentum $p_T$ at (a) {\it HERA--}$\vec{N}$
and (b) RHIC ($\sqrt{s} = 200$~GeV).  
 The solid (dashed) line corresponds
to set A (B) of the NLO GS parameterization \protect \cite{GSnew}
 and the dash-dotted line to the LO set A \protect \cite{GSnew}.
The dotted line represents the expected asymmetry calculated 
only in the CSM for the NLO set A \protect \cite{GSnew}.
The projected statistical sensitivities are also shown. 
The figures are taken from ref.~\cite{nowaktkabl}.
 }
\label{chi2avto}
\end{figure}

Recently, double spin asymmetries in $P$-wave charmonium production at 
{\it non-zero} transverse momentum ($p_t > 1.5$ GeV) were calculated
\cite{nowaktkabl} taking into account both possible mechanisms,
CSM and COM.
The asymmetries for the production
of $\chi_{c1}$ and $\chi_{c2}$ states exhibit different signs in the
transverse momentum range $1.5<p_t<3$ GeV.
The transverse momentum dependence of
$A_{LL}^{\chi_{c2}}$ is shown, as an example, in fig.~\ref{chi2avto}
for certain parameterizations of the polarized gluon distribution.
Additionally, projected statistical errors were calculated for both
{\it  HERA--}$\vec{N}$ and RHIC.

A measurement of $A_{LL}^{\chi_{c2}}$ for $1.5 < p_t < 2.5$ GeV
would clearly identify a polarized gluon distribution if it is
sufficiently large at $x_{gluon} \simeq 0.1$ (see fig.~\ref{chi2avto}a).
Additionally, a differentiation between color
singlet and color octet contributions may be possible in this case.

With increasing c.m.s. energy the discrimination power of $\chi_{c1,c2}$ 
production decreases, as can be seen from fig.~\ref{chi2avto}b
showing the double spin asymmetry at $\sqrt{s}=200$ GeV. With the
presently envisaged integrated {\it RHIC} luminosity of 320~pb$^{-1}$
assuming $100\%$ efficiency and $P_B P_T\simeq0.5$ it will 
be very hard to draw any conclusion in the discussed context.

Very recently $\chi_{c2}$ production at small transverse momenta was
considered based upon the ideas of QCD multipole
expansion \cite{chi2jaffe}.
It is argued that a measurement of the angular distribution in the
radiative decay of $\chi_{c2}$ produced in {\it unpolarized} proton-proton
collisions is supposed to yield the ratio 
between the amplitudes of color octet and color singlet states.
Knowing this ratio a measurement of the angular
distribution of the double spin asymmetry in {\it polarized}
proton-proton is then supposed to allow access to $\Delta G \over G$.

It seems appropriate here to underline that the inclusive photon and
charmonium final states discussed above correspond to
different ranges of validity of 
pQCD. When measuring photon production with transverse momenta
between 2~GeV and 6~GeV the lower momentum region can not be
considered as a safe working ground for pQCD. In contrast, charmonium
production is safe even at vanishing transverse momentum since the
large mass of the charm quark guarantees the necessary hard scale.  \\

\vspace*{-5mm}

\subsection{ Photon or $J/\psi$ Production Associated with Jets}

\vspace{1mm}
\noindent
In contrast to inclusive production the final state 
`photon ($J/\psi$) plus jet' offers {\it direct} access to the
polarized gluon distribution. 
Since the underlying partonic subprocess is of the
type $2 \rightarrow 2$, the kinematics of the back-to-back parton
emission is completely known, at least in principle, if the emerging
products on the hadron level are fully detected. This is an essential
advantage compared to inclusive production discussed in the
previous section.

%
{\bf Photon plus jet production.}
The complete kinematics of the underlying hard 2$\rightarrow$2 subprocess
allows to establish a simple relation between 
$A_{LL}^{\; \gamma+jet}$ and $\Delta G \over G$ \cite{desy96-04}:
\begin{eqnarray}
\label{DelGtoGphot}
{\Delta G \over G}(x_{gluon}) = {A_{LL}^{\; \gamma+jet} \over 
        {A_{DIS} \cdot \hat a_{LL} }},
\end{eqnarray}
where the partonic level asymmetry $\hat a_{LL}$ and $A_{DIS} = g_1/F_1$
should be taken at appropriate values of the kinematical variables
calculated from the measured kinematics of the registered photon and
jet.

\begin{figure}[hbt]
\begin{center}
\epsfig{file=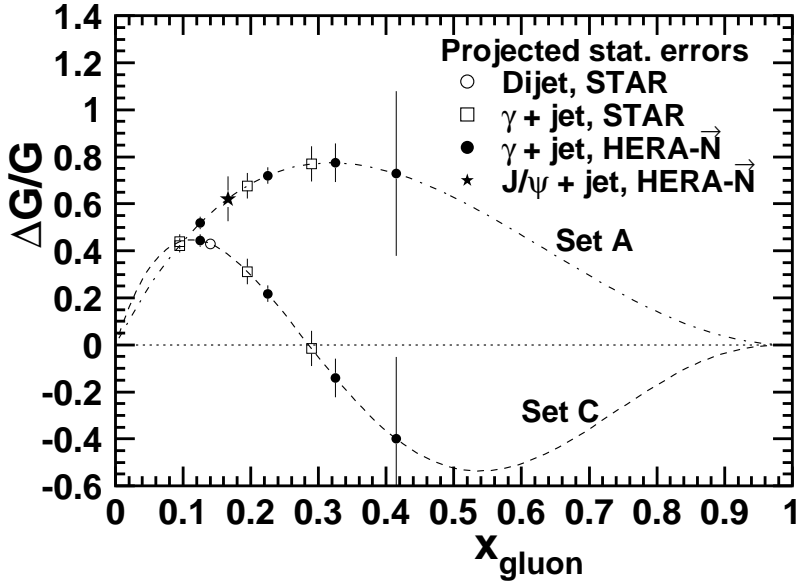,width=12cm}
\caption{
\it Typical predictions for the polarized gluon
       distribution (LO calculations
       from ref. \protect \cite{GehrStir}) confronted
       to the projected statistical errors expected for
       {\it HERA--}$\vec{N}$ and RHIC experiments.  }
\label{gamjet}
\end{center}
\end{figure}

This approach was used in ref.~\cite{desy96_128} to estimate
the projected statistical sensitivity of
{\it HERA--}$\vec{N}$ for a ${\Delta G \over G}(x_{gluon})$ measurement.
The corresponding results, including the acceptance of a
possible detector, are shown in fig.~\ref{gamjet} vs. $x_{g}$ in
conjunction with predicted errors for {\it STAR} running at RHIC at
200 GeV c.m. energy \cite{yok1}.
The errors demonstrate clearly that in the region
$0.1 \leq x_{gluon} \leq 0.4$ a significant result from photon plus
jet production can be expected from {\it HERA--}$\vec{N}$.
As can be seen, the measurement
of $\Delta G \over G $ from photon plus jet production
in doubly polarized nucleon-nucleon collisions at HERA can
presumably be performed with an accuracy that is competitive
to RHIC.  

It seems here appropriate to underline that the interesting
$x_{gluon}$-range (0.1~...~0.3) will be accessed with {\it STAR} at 200~GeV in
the deep perturbative transverse momentum region (10~...~30~GeV),
whereas at {\it HERA--}$\vec{N}$ the transverse momenta lie in the
pQCD onset region (2~...~6~GeV); hence both approaches are complementary
with respect to the validity of pQCD. 

It is important to note that the {\it HERA--}$\vec{N}$
fixed-target kinematics causes additional problems for jet
reconstruction in comparison to a collider experiment
(for more details see ref.~\cite{desy96-04}). 
A forward oriented detector with good granularity down to scattering 
angles of $10 \div 20$ mrad is required. 
From preliminary Monte Carlo studies \cite{desy96-04} it was
seen that the number of photon events accompanied by a successfully
reconstructed jet decreases considerably for lower values
of $p_T$ and, correspondingly, of $x_{gluon}$. Appropriate preliminary
jet reconstruction efficiencies were included to arrive at realistic
projected error bars for the {\it HERA--}$\vec{N}$ points in 
fig.~\ref{gamjet}.   \\

{\bf $J/\psi$ plus jet production.}
Once the mechanism of $J/\psi$ production is
established (see discussion in the previous section),
a measurement of the double spin asymmetry $A_{LL}^{J/\psi+jet}$
would allow to access the polarized gluon distribution
function directly, in a similar manner as in case of photon 
plus jet production discussed above. The absolute statistical 
error of ${\Delta G \over G}(x_{gluon})$ can be expressed as:
\vspace{-1em}
\begin{eqnarray}
\label{err1}
\delta [{\Delta G \over G}(x_{gluon})]=
{\delta A_{LL}^{J/\psi+jet} \over {[\Delta G/G] \cdot \hat a_{LL}}} .
\end{eqnarray}

In the case of $J/\psi$ plus jet
production the cross section decreases more rapidly with $x_{gluon}$.
Following the same principle of analysis as ref. \cite{desy96-04},
it turns out that the measurement of  $\Delta G \over G$ in $J/\psi$
plus jet production is feasible only for $x_{gluon}~=~0.1~\div~0.2$,
i.e. for $J/\psi$ transverse momenta of about 2.5 GeV/c. This
prediction is shown as an additional entry in fig.~\ref{gamjet}. 
Although it is only a single point, this is a very important
measurement, because the lowest point from photon plus jet
production is obtained for small transverse momentum where
perturbative QCD is not expected to give very reliable predictions. It
is worth noting that the nature of the gluon-gluon subprocess has
similar consequences for process of jet plus jet production at RHIC.
The prediction \cite{yok1} consists of only one point at a comparable
value of $x_{gluon}$ (see fig.~\ref{gamjet}). 

\vspace*{-2mm}

\subsection{Other Processes}

\vspace*{-1mm}

As just mentioned, also the possibility to extract $\Delta G / G$ 
from inclusive {\bf two-jet production} may be considered. 
Several hard subprocesses
($gg$, $gq$, $qq$ scattering) contribute to the two-jet final state
and a good knowledge of polarized quark distributions would be 
essential. Note that at {\it HERA--}$\vec{N}$ the
above mentioned additional difficulty of jet reconstruction
must be addressed. A potential solution of the problem consists in 
replacing the reconstruction of the two jets by the
registration of two correlated high-$p_T$ hadrons opposite in azimuth. 

Another, still unexplored possibility is {\bf open charm production}.
Interesting results might be obtained using a high $p_T$ single muon 
or electron-muon pairs from charm decays as a tag.
Both possibilities still need  careful investigation.\\

\vspace*{-3mm}

\section{Anti-Quark Helicity Distributions $\Delta \bar{q}(x)$ }
\label{dylong}

The production of Drell-Yan pairs in nucleon-nucleon collisions
proceeds via quark-antiquark annihilation, 
$q(x_1) + \bar q (x_2) \rightarrow \gamma ^* \rightarrow l^+ l^-$.
The {\it longitudinal} double spin asymmetry turns out to be well
suited to extract the polarized light sea-quark distribution; the
asymmetry can be written as
\begin{eqnarray}
\label{DYll}
 A_{LL}^{DY} = 
          {{ \sum_f \: e^2_f \: \Delta q_f(x_1) \: \Delta
           \bar q_f(x_2) \: d \, \Delta \hat \sigma(x_1,x_2)+[x_1
           \leftrightarrow x_2]}
        \over
          {\sum_f \: e^2_f \: q_f(x_1) \: \bar q_f(x_2) \:
          d \, \hat \sigma(x_1,x_2)+[x_1 \leftrightarrow x_2]}}.
\end{eqnarray}
Note that the parton level asymmetry in Drell-Yan production is maximal, 
$\hat a_{LL} = -1$.

\begin{figure}
\begin{center}
\epsfig{file=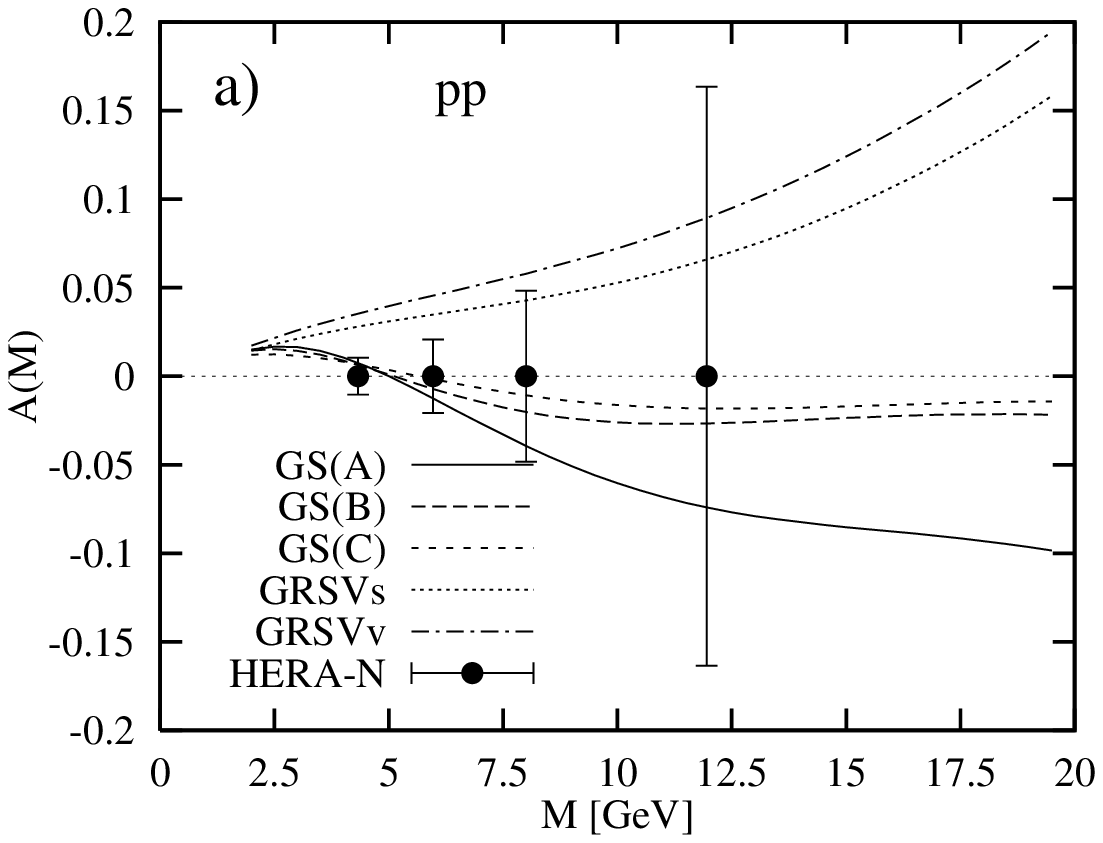,width=7.3cm}
\epsfig{file=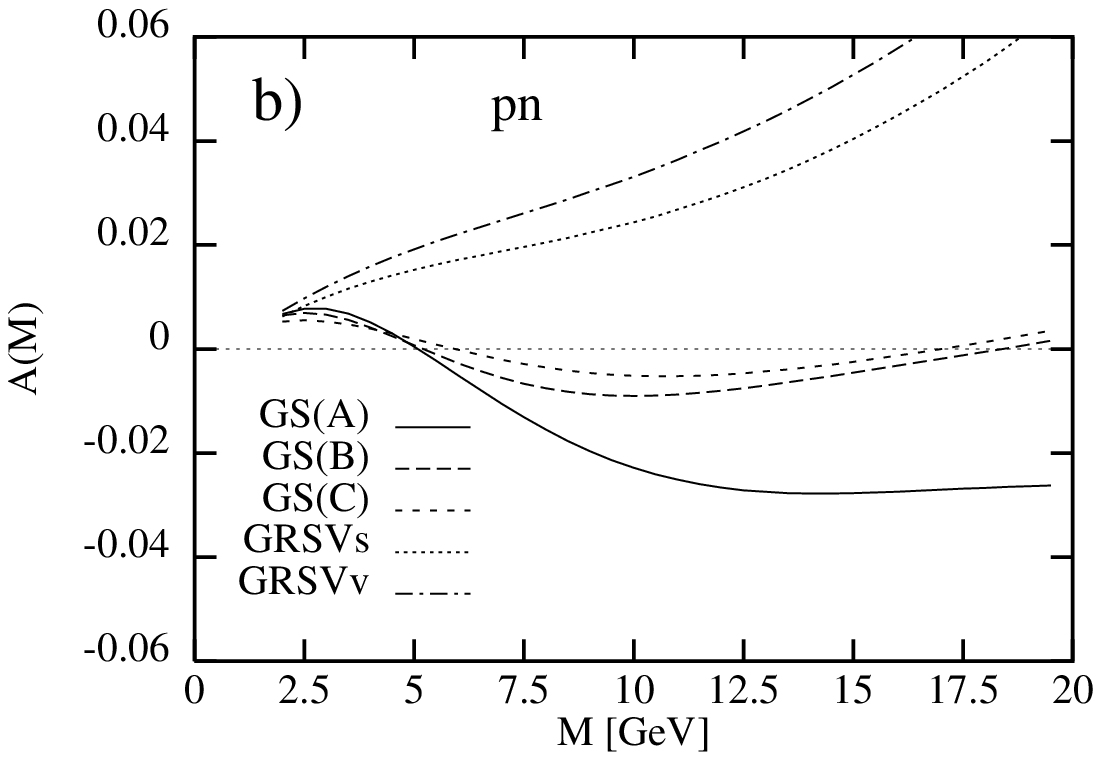,width=8.2cm}
\caption{
\it Longitudinal double spin asymmetries in the polarized
    Drell-Yan process \protect\cite{DYGehrStir} for (a) pp and 
    (b) pn collisions ($\sqrt s = 40$~GeV) confronted to the projected
       {\it HERA--}$\vec{N}$ statistical errors.}
\label{dygehr}
\end{center}
\vspace*{-4mm}
\end{figure}

The prospects for the $A_{LL}^{DY}$ measurement at 
{\it HERA--}$\vec{N}$ were calculated \cite{DYGehrStir} in
next-to-leading order QCD. The spread of the predictions (see
fig.~\ref{dygehr}a,b) reflects the insufficient present knowledge on
the polarized sea quark distributions in the region $x > 0.1$; not
even the sign of the asymmetry is predicted at large mass $M$ of the pair.
The asymmetry is obtained from the weighted sum of $\Delta \bar u$ and 
$\Delta \bar d$ quarks; the strange quark contribution is assumed to
be small. In the case of a proton target the weight of $\Delta \bar u$ is
higher than that of $\Delta \bar d$ due to its abundance in the proton
(and the electric charge). Hence the asymmetry expected in $pp$
collisions (fig.~\ref{dygehr}a) provides mainly information on $\Delta
\bar u$, i.e. on the sea $u$ quark polarization. The flavor
contributions are different for $pn$ collisions; this results in a
much smaller asymmetry (fig.~\ref{dygehr}b) than in the $pp$
case (note the different vertical scales in fig.'s~\ref{dygehr}a and b). 
The projected statistical uncertainties for {\it HERA--}$\vec{N}$, as
shown in fig.~\ref{dygehr}a, have to be enlarged by a factor of
$\sqrt{2}$ since preliminary acceptance calculations  based upon a
{\it HERA--B} like detector resulted in a value of about 0.5
\cite{KoNoZtn97}. Then, with a proton target, a statistically
significant measurement of $A_{LL}^{DY}$ would require an integrated
luminosity about twice as large as the anticipated 240~pb$^{-1}$, which
appears not unrealistic. Suggestions to extract more 
detailed physics information by measuring the differential lepton pair
distributions in dependence on $x_F$ or $\eta$
\cite{DYGehrStir,DYGehr} would require even higher luminosities. This
applies as well to studies of Drell-Yan production on a neutron target
where asymmetries smaller by about a factor of three compared to the
proton case had to be probed, as can be
seen by comparing the scales of fig.'s~\ref{dygehr}a and b.

\section{Transversity distribution $\delta q(x)$}

\begin{figure}
\vspace*{-5mm}
\begin{minipage}[c]{8cm}
\centering
\epsfig{file=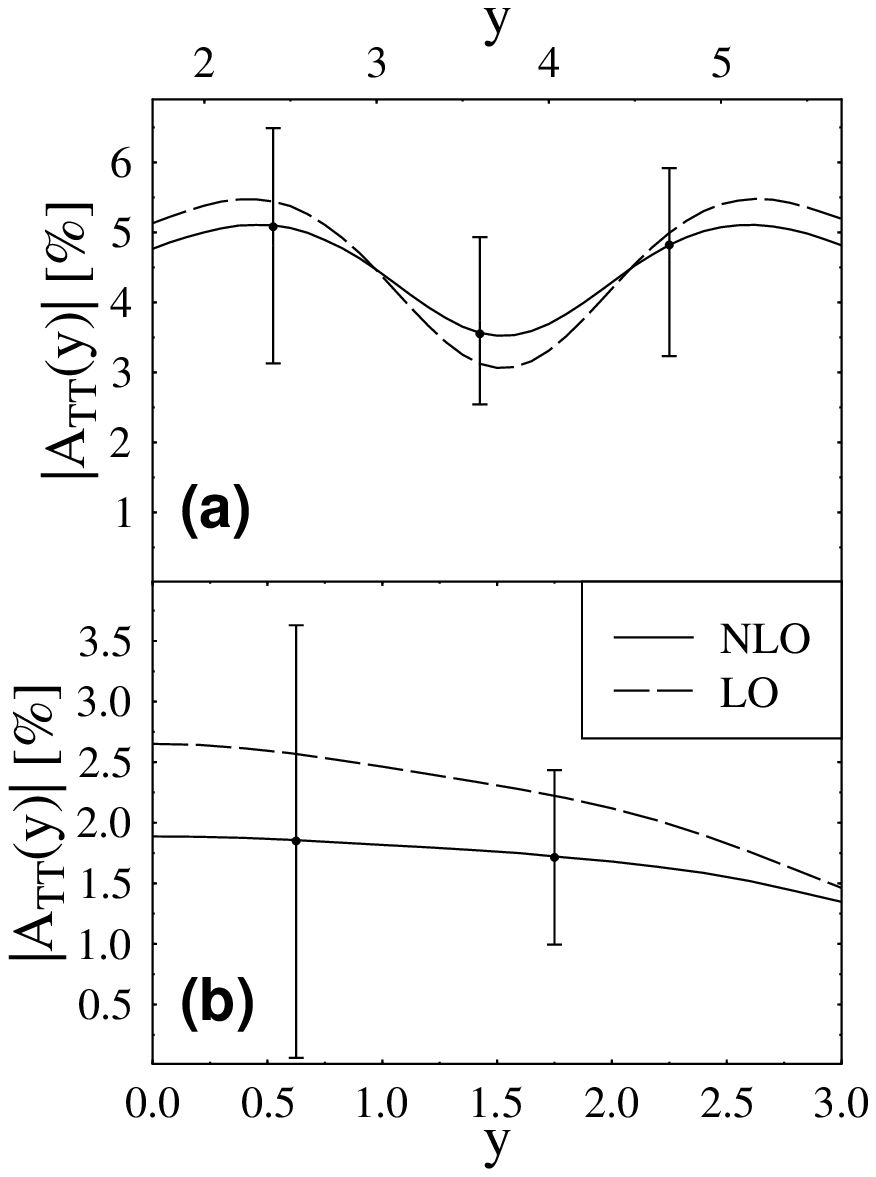,width=6.0cm}
\end{minipage}
\begin{minipage}[c]{8cm}
\centering
\epsfig{file=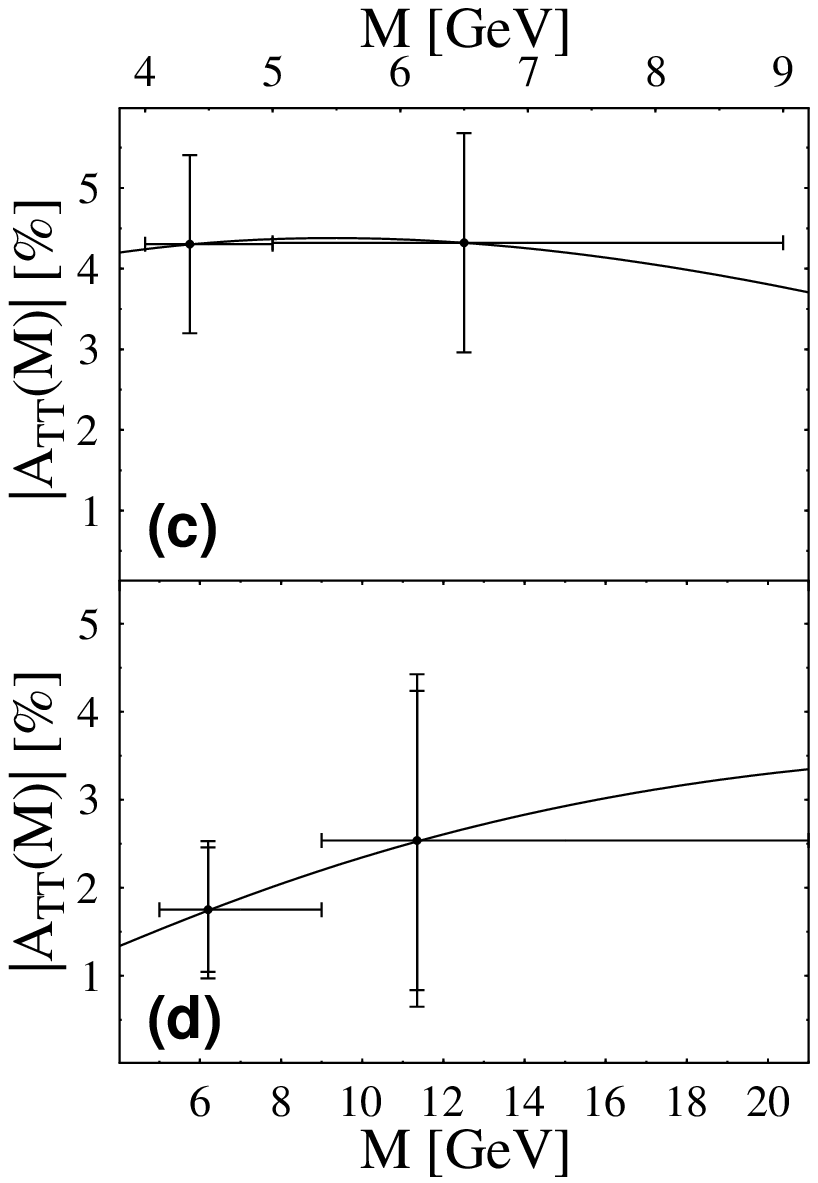,width=6.0cm}
\vspace*{+2mm}
\end{minipage}
\caption{
\it Predictions \cite{martin} for the rapidity dependence of
$A_{TT}$ for (a) {\it HERA--}$\vec{N}$ ($E_{\rm beam}=820$~GeV, $M=4-9$~GeV) 
and (b) PHENIX at RHIC ($\sqrt{S}=200$~GeV, $M=5-9$~GeV).
Predictions \cite{martin} of $A_{TT}(M)$ for 
(c) {\it HERA--}$\vec{N}$ and (d)  RHIC. 
}
\label{figmartin}
\end{figure}

Drell-Yan pair production with {\it transverse} polarization of
{\it both} beam and target can provide a qualitatively new insight
into the spin structure of the nucleon by measuring the third twist-2
quark distribution function (quark transversity
distribution, $\delta q(x)$) which is absolutely unknown
at present. It essentially describes the fraction of transverse
polarization of the proton carried by its quarks. In inclusive
DIS a contribution from this function is suppressed by a factor 
containing the quark mass
whereas it is in principle accessible in semi-inclusive DIS. \\
The asymmetry $A^{DY}_{TT}$ can
be written in the form \cite{JaffeJi2}
\begin{equation}
\label{DYTT}
A^{DY}_{TT} = {{\sin^2 \theta \cos{2 \phi}} \over {1 + \cos^2 \theta}}
\quad {{\sum_f e^2_f \:\delta q_f(x_1) \, \delta \bar q_f(x_2) + 
[x_1 \leftrightarrow x_2]} \over 
{\sum_f e^2_f \: q_f(x_1) \, \bar q_f(x_2) + 
[x_1 \leftrightarrow x_2]} },
\end{equation}
where $\theta$ is the polar angle of one of the leptons in the 
virtual photon
rest frame and $\phi$ is the angle between the direction of the
nucleon polarization and the normal to the dilepton decay plane. The
asymmetry vanishes on integration over the azimuthal angle.   \\
Due to the lack of any information on the transversity distribution,
the largest possible value of the asymmetry was estimated 
\cite{martin}, based on a saturation of Soffer's inequality 
\cite{soffer95}. The resulting LO and NLO asymmetries are presented in
fig.~\ref{figmartin}a and b for {\it HERA--}$\vec{N}$ and RHIC
kinematics, respectively. Fig.~\ref{figmartin} also shows the 
projected statistical errors for a measurement of $A_{TT}$ including
detector acceptance effects. The maximal value of $A_{TT}$
at an invariant mass of $M=4$ GeV was found to be approximately 0.04
with a projected statistical error of about 0.01. The expected value
of the asymmetry at RHIC energies is smaller as is the expected
statistical significance of the measurement.

The {\bf longitudinal-transverse} double spin asymmetry,
$A^{DY}_{LT}$, is sensitive to the transversity distribution, although 
the asymmetry receives additionally contributions from the twist-3
distributions $g_T$ and $h_L$. However, calculations \cite{kanazawa}
show that $A^{DY}_{LT}$ is clearly expected to be much
smaller than $A^{DY}_{TT}$ while the absolute predictions still 
suffer from the limited accuracy of the involved parton distributions.

{\bf Two-meson production.}
Recently it has been proposed to probe the nucleon's transversity
distribution through the final state interaction between two mesons 
($\pi^+ \pi^-, \pi K$, or $K \bar{K}$) inclusively produced 
on a transversely polarised nucleon, 
$p~+~p^{\uparrow}\rightarrow~ \pi^+~+\pi^-~+~X$ \cite{JJT1,tang}. 
It has been shown that the interference effect
between the s- and p-wave of the two-meson system  around the $\rho$
(for pions), $K^*$ (for $\pi K$), or $\phi$ (for kaons) provides an
asymmetry which may be sensitive to the quark transversity
distribution in the nucleon.  The asymmetry is a function
of the angle of the normal of the two-pion plane $\vec{k}_+ \times
\vec{k}_-$ with respect to the polarization vector $\vec{S}_\perp$
of the nucleon. \\
An asymmetry as large as
0.15 has been predicted \cite{tang} for jet energies of about 100 GeV at
RHIC. No estimates are available yet for {\it HERA--}$\vec N$ where a
corresponding measurement could be done already in 'Phase-I', i.e. with
an {\it un}polarized beam.

\newpage

\section{Polarized Fragmentation Functions}

When studying the spin transfer from a {\it longitudinally} polarized
nucleon to the $\Lambda$ hyperon in the process $p\vec p \rightarrow
\vec \Lambda X$ at large $\Lambda$ transverse momentum,
the relevant spin asymmetry is defined as
\begin{equation}
\label{asylambda}
 A^{\Lambda} = 
{{d \Delta \sigma^{p\vec p \rightarrow \vec{\Lambda}X} / d \eta}
\over {d \sigma^{p p \rightarrow \Lambda X} / d \eta}}.
\end{equation}
Here $d \sigma^{p p \rightarrow \Lambda X} / d \eta$ is the
unpolarized cross-section and 
$d \Delta \sigma^{p\vec p \rightarrow \vec{\Lambda}X} / d \eta$
is given by
\begin{equation}
\label{dsiglambda}
{d \Delta \sigma^{p\vec p \rightarrow \vec{\Lambda}X} / d \eta} =
{d \sigma^{p p_+ \rightarrow \Lambda_+ X} / d \eta} -
{d \sigma^{p p_- \rightarrow \Lambda_+ X} / d \eta} .
\end{equation}
The subscripts $"+","-"$ denote particle helicities. 

\begin{figure}[hbt]
\begin{center}
\epsfig{file=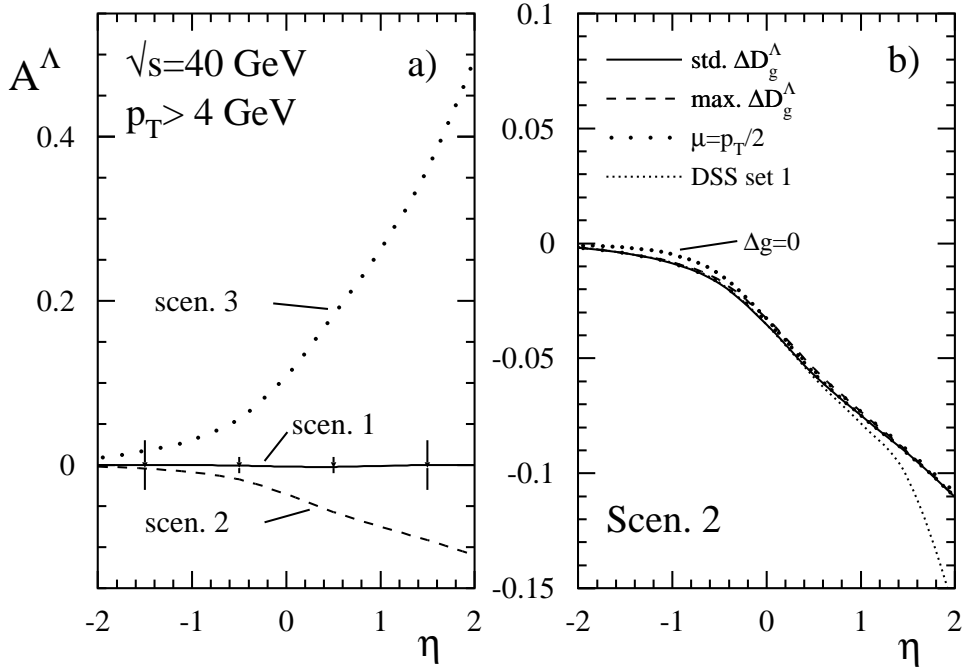,width=13cm}
\caption{ \it The asymmetry $A^\Lambda$ as defined in 
Eq.~(\ref{asylambda}) as a function of $\Lambda$ rapidity 
at {\it HERA--}$\vec{N}$ energy for various sets of spin-dependent
fragmentation functions together with the projected statistical
errors. A discussion of the different scenarios and curves can be
found in ref.~\cite{florian2}. }
\label{figflorian}
\end{center}
\end{figure}

An attempt to determine the spin-dependent $\Lambda$ fragmentation
functions showed \cite{florian1} that the available LEP data cannot
sufficiently constrain the valence fragmentation functions
for all flavors. Rather different scenarios adopted for the input
valence distributions appear to describe the data equally well.
The LO calculation \cite{florian2} shows that a measurement 
of the asymmetry~(\ref{asylambda}) at {\it HERA--}$\vec{N}$
may allow to discriminate among different possible alternatives 
(see fig.~\ref{figflorian}). \\

\newpage

\section{Deuteron and Helium Targets}

\noindent

A fixed-target experiment like {\it HERA--}$\vec{N}$ offers a unique
possibility to study polarized $pn$ and $pd$ collisions which are harder
or even impossible to realize in collider experiments at RHIC.
Using a polarized $^3He$ target would allow to investigate 
polarized $pn$ collisions and in particular a measurement of the
polarized 
sea quark distributions via Drell-Yan pair production 
(see section \ref{dylong}). Moreover, it is very likely that a
fixed-target environment offers the only chance to investigate $pd$
collisions with longitudinally polarized deuterons since it presently
appears rather difficult to attain longitudinal polarization for 
deuterons in a collider due to their small magnetic moment.

Compared to the proton target, the deuteron, as a spin-1
hadron, has additional twist-2 parton distributions, in particular 
a tensor-polarized quark distribution, $b_1(x)$. 
It could be measured using Drell-Yan pair production in
scattering {\it unpolarized} protons on the {\it tensor-polarized}
deuteron. The corresponding asymmetry can be written as \cite{kumano} 
\begin{eqnarray}
\label{DYtenzor}
 A_{UQ_0}^{DY} = 
          {{ \sum_f \: e^2_f \: [ q_f(x_1) \: \bar b_{1f}(x_2) \: + 
           \: \bar q_f(x_2) \: b_{1f}(x1) ] } 
        \over
          {\sum_f \: e^2_f \: [ q_f(x_1) \: \bar q_f(x_2) \: +
           \: \bar q_f(x_2) \: q_f(x_1) ] }}.
\end{eqnarray}
No quantitative estimates for the asymmetry are available yet.

\section{Summary and Conclusions}

\noindent

The physics case for a possible fixed target polarized nucleon-nucleon 
collision experiment utilizing an internal target in the 820 GeV HERA
proton beam has been presented. A wide spectrum of nucleon spin
structure problems could be investigated. Single (transverse) spin
asymmetries, accessible already with the existing unpolarized 
beam, are found to be a powerful tool to study the nature
and physical origin of higher twist effects and a possible manifestation of
non-perturbative dynamics. Their measurement requires a sufficiently
large $p_T$-range; {\it HERA--}$\vec{N}$ would be able to provide data up
to $p_T=$10~GeV/c in the central region and up to 5$\div$6~GeV/c in
the fragmentation region of the polarized nucleon.
When measuring the polarized gluon distribution through double spin 
asymmetries in {\it photon (plus jet)} and {\it $J/\psi$ (plus jet)} 
production -- requiring a polarized HERA proton beam -- the projected 
statistical accuracies are found to be comparable to those predicted for the 
spin physics program at RHIC. 
Although both approaches explore the same $x_{gluon}$ range they are
complementary due to the different accessible $p_T$ ranges.
A measurement of
Drell-Yan pair production with both beam and target longitudinally
polarized can improve our knowledge on the polarized light sea quark 
distributions. A study of double transverse 
and/or longitudinal-transverse Drell-Yan spin asymmetries as well as a
study of the single (transverse) spin asymmetry in inclusive two-pion
production may open an access to the quark transversity distribution(s). 
A study of the spin transfer in the reaction $p\vec p \rightarrow \vec
\Lambda X$ is capable of providing essential constraints on the
polarized $\Lambda$ hyperon fragmentation functions.
The existence of a polarized internal gas target in 
{\it HERA--}$\vec{N}$ would 
allow to study also polarized $pn$ and $pA( D, ^3He, ...)$
collisions which are harder or even impossible to realize in
collider experiments at RHIC.
In addition, there is a potential to obtain significant results on
the long-standing unexplained spin asymmetries in elastic scattering.

\section*{Acknowledgments}

\noindent
We thank R. Kaiser for the careful reading of the manuscript.

\end{document}